\documentclass[a4paper,11pt]{article}
\pdfoutput=1
\usepackage{jcappub}
\usepackage{aas_macros}
\usepackage{subcaption}

\usepackage{graphicx}
\makeatletter
\let\orig@Gin@extensions\Gin@extensions 
\def\Gin@extensions{.pdf,\orig@Gin@extensions} 
\makeatother

\usepackage[normalem]{ulem}

\newcommand{\MZ}[1]{\textcolor{blue}{MZ: #1}}
\def\bi#1{\hbox{\boldmath{$#1$}}}

\newcommand{\loss}{\mathcal{L}}
\newcommand{\be}{\begin{equation}}
\newcommand{\ee}{\end{equation}}
\newcommand{\mpc}{\, {\rm Mpc}}

\newcommand{\hmpc}{\, h^{-1} \mpc}

\title{Exploring the posterior surface of the large scale structure reconstruction}

\author[a,c]{Yu Feng,}
\author[a,b,c]{Uro{\v s} Seljak}
\author[d]{\& Matias Zaldarriaga}

\affiliation[a]{Department of Physics, University of California, Berkeley, CA 94720, USA}
\affiliation[b]{Lawrence Berkeley National Laboratory, Berkeley, CA 94720, USA}
\affiliation[c]{Berkeley Center for Cosmological Physics, University of California, Berkeley, CA 94720, USA}
\affiliation[d]{Institute for Advanced Study, Princeton, NJ 08540, USA}

\emailAdd{yfeng1@berkeley.edu}
\emailAdd{useljak@berkeley.edu}
\emailAdd{matiasz@ias.edu}
\abstract{The large scale structure (LSS) of the universe is generated by the linear density gaussian modes,
which are evolved into the observed nonlinear LSS.
Given a set of data the posterior surface of the modes is convex in the linear regime,
leading to a unique global maximum (MAP), 
but this is no longer guaranteed in the nonlinear regime.
In this paper we investigate the 
nature of posterior surface using the recently developed MAP reconstruction method,
with a simplified but realistic N-body simulation as the forward model.
The reconstruction method uses optimization with analytic gradients from 
back-propagation through the simulation.
For low noise cases we recover the initial conditions well into the nonlinear regime ($k\sim 1$ h/Mpc)
nearly perfectly.
We show that the large scale modes can be recovered more precisely than the 
linear expectation, which we argue is a consequence of nonlinear mode coupling. 
For noise levels achievable with current and planned LSS surveys the reconstruction cannot recover 
very small scales due to noise.
We see some evidence of non-convexity, specially for smaller scales where 
the non-injective nature of the mappings: several very different initial conditions leading
to the same near perfect final data reconstruction. 
We investigate the nature of these phenomena further using a 1-d 
toy gravity model, where many well separated local maximas
are found to have identical data likelihood but differ in the prior.
We also show that in 1-d the prior favors some solutions over the true solution,
though no clear evidence of these in 3-d.
Our main conclusion is that on very small scales and for a very low noise 
the posterior surface is multi-modal and the global maximum may be unreachable
with standard methods, 
while for realistic noise levels in the context of the current and next generation
LSS surveys MAP optimization method is likely to be nearly optimal.}

\begin{document}
\maketitle

\section{Introduction}

The large scale structure (LSS) of the universe can be observed through a 
number of methods. Among these are galaxy positions in redshift space or 
in angular position in the sky, weak lensing maps of projected mass density, 
galaxy clusters observed through a number of techniques (optical, X-ray, SZ...), 
Lyman alpha forest, 21cm in emission or absorption etc. A common theme in LSS 
is that these observations can be connected to the underlying 
distribution of matter, which is dominated by the dark matter. This connection 
is not simple, and given sparse and noisy sampling of the typical LSS 
dataset we can only incompletely reconstruct the nonlinear matter distribution. 
Nevertheless, it is believed that the observed properties of LSS are to a 
large extent determined by the dark matter distribution, and 
even if baryonic effects must be included in the complete physical model, 
they can to a large extent be connected to the dark matter properties, such as density, 
velocity dispersion, or tidal tensor.

Assuming that we are able to use the observations to determine a noisy version of the nonlinear dark matter density distribution, we can ask how can this be related to the initial conditions, where the modes are Gaussian distributed 
under the 
assumption that the physical process that generates them (e.g. inflation) is Gaussian. If one could perform this operation 
perfectly one would eliminate all the non-Gaussian features present in the 
final, nonlinear dark matter distribution. In terms of information transfer one 
would convert the information in the higher order non-Gaussian statistics into 
the information purely encoded in the two-point function statistics of initial 
density field. This would be hugely beneficial as there is currently no 
known method that can optimally extract all the information present in the nonlinear 
data. 

There have been a few attempts to perform such a reconstruction in the 
literature: the methods range from the simplified dynamics models \cite{EsensteinEtAl07,SchmittfullEtAl17,ZhuEtAl17}, 
to the Bayesian reconstruction using simplified dynamics or 
fully nonlinear dynamical model \cite{2008MNRAS.389..497K,2010MNRAS.409..355J,JascheWandelt13,2013ApJ...779...15J,WangMoEtAl14,JascheLeclercqEtAl15,SeljakEtAl17}. 
Typically we expect the reconstruction to fail in the presence of noise and 
incomplete data, and because of this one either 
tries to find maximum a posteriori (MAP) solution \cite{WangMoEtAl14,SeljakEtAl17}, or one
makes sample realizations using Gibbs or Hamiltonian sampling \cite{2013ApJ...779...15J,JascheLeclercqEtAl15}. The latter 
approach is expensive and not needed 
in the linear regime, where the posterior surface is convex and  
MAP is unique for any level of noise and coverage. In the 
nonlinear regime the posterior may become multi-modal, but 
this does not mean that the methods must fail. For example, 
most of the posterior mass may still be
concentrated around a single point with approximately a
Gaussian distribution around it. Alternatively, we may have several posterior maxima
which are all equivalent in terms of their information content (defined more precisely 
below). 
In this paper we explore the validity of the MAP approach in the nonlinear regime, where the 
posterior surface is no longer guaranteed to be convex. 
We will explore posterior surface using nonlinear dynamics with finite step 
N-body simulations using FastPM scheme \cite{FengChuEtAl16}. 

Ultimately, our goal is not to obtain MAP at a fixed fiducial 
power spectrum, but to create an optimal estimator of 
the power spectrum itself, and the 
cosmological parameters encoded in it (cosmological parameters are also in 
the parameters that control the mapping from the initial to the final density, which 
we will fix for the purpose of this paper, as well as non-standard parameters such as primordial non-Gaussianity). 
In the formalism of \cite{SeljakEtAl17}, the starting point for finding the optimal power spectrum 
estimator is the MAP evaluated at 
a fiducial power spectrum. If the posterior surface is convex and can be written as a 
multi-variate Gaussian around the MAP one can perform analytic Gaussian
marginalization over the modes, after which one is left with the posterior of the 
power spectrum given the data. The data come into this expression entirely in the 
form of the MAP solution itself. One can explore this posterior surface by 
performing Newton-Raphson's expansion, which leads to the peak posterior 
values of the power spectrum bandpowers and the associated covariance matrix that 
can both be expressed in terms of the MAP values. An additional term, 
a posterior volume, also needs to be evaluated, and stochastic integration using
simulations has been proposed in \cite{SeljakEtAl17} as a way to evaluate it due to the high dimensionality
of the problem. 

This procedure is justified if the 
posterior surface is convex, or if one can approximate most of the posterior 
volume by the values around MAP. 
In this approach
MAP plays a central role in the optimal parameter determination, and the question we want to 
address is whether this is still valid in the nonlinear regime. 
One also expects that if the 
MAP solution is unique, one should be able to find it in the limit of low 
noise and complete coverage. This is just a statement that if we are able to 
obtain the true linear solution from which the data have been generated, it 
would contain all the information, so we would just evaluate its power 
spectrum and determine all the information from it. In this paper we ask the question of posterior 
surface of modes at a fixed power spectrum, which we will for simplicity
fix at the true power spectrum. We will ask whether different starting points lead to the 
same MAP, or whether there exist different local maxima in the posterior surface that can 
complicate the interpretation. 

The plan for this paper is the following. In section \ref{sec2} we present
a series of numerical studies of reconstruction of initial conditions, 
using a full N-body simulation for the forward dynamical model. We vary 
various parameters, including noise, starting points for optimization, 
random seed etc., to explore the nature of solutions.  
In the following section \ref{sec3} we present a toy model that attempts 
to qualitatively explain the main numerical 
results found in previous section. This is followed by a discussion and 
conclusions section \ref{sec4}. 
\section{Numerical Method}
\label{sec:numerical}
\subsection{Objective function}
The setup we consider is the following \cite{SeljakEtAl17}: 
we measure dark matter density sampled on a 3-d grid,
arranged into a vector 
$\bi{d}=\{d(\bi{r}_i)\}(i=1,...,N)$. 
Each measurement consists of a signal and a noise 
contribution, $\bi{d}=\bi{f(s)}+\bi{n}$, where noise is 
assumed to be uncorrelated with the signal. 
Here $\bi{f(s)}$ is 
a nonlinear mapping from the initial linear density modes to the final model 
prediction of the observation,
and $\bi{s}=\{s_j\}(j=1,...,M)$ 
are the underlying initial density mode coefficients that we wish to estimate.
The modes $\bi{s}$ are latent variables with a prior 
parametrized as a multivariate Gaussian, 
with covariance matrix
$\bi{S}=\langle \bi{ss}^{\dag}\rangle$, which is assumed to be diagonal 
in Fourier space. We will assume the power spectrum is known, although in 
practice it can only be known to the extent that the data allow it. 
The conditional posterior
probability for $\bi{s}$ given the data and prior is
\be
P(\bi{s} \vert \bi{d})= (2\pi)^{-(M+N)/2} \det(\bi{SN})^{-1/2}\exp\left(-{\frac{1}{2}}
\left\{\bi{s}^{\dag} \bi{S}^{-1}\bi{s}+\left[\bi{d-f(s)}\right]^{\dag} \bi{N}^{-1}\left[\bi{d-f(s)}\right]\right\}\right).
\label{lik}
\ee

The MAP solution of the initial modes can be obtained by maximizing
the posterior of $\bi{s}$ at a fixed $\bi{S}$ and $\bi{N}$, 
which is equivalent to minimizing the loss function $\mathcal{L}(\bi{s})$
\be
\loss(\bi{s})\equiv -2\log P(\bi{s} \vert \bi{d})=\bi{s}^{\dag} \bi{S}^{-1}\bi{s}+[\bi{d-f(s)}]^{\dag} \bi{N}^{-1}[\bi{d-f(s)]},
\label{chi2_eq}
\ee
with respect to all parameters $\bi{s}$. We have dropped the terms that 
do not depend on $\bi{s}$. If we assume noise is 
diagonal in real space, $N_{ij}=\sigma^2\delta_{ij}$, we obtain an objective function as two sums of squares, over all the modes $s_k$ and over all the data points $d_i$,
\begin{equation}
\loss({\bi s}) = \sum_k \frac{|s_k|^2}{S_{kk}} + \sum_i \left(\frac{d_i-f_i(\bi{s}) }{\sigma}\right)^2 = \loss_s + \loss_d,
\label{eq:obj}
\end{equation}
where we labeled the the prior term (first term) $\loss_s$ and the data residual term (second term) $\loss_d$. A maximum in the posterior implies a minimum in the loss function: we will use both 
languages here. We refer to a global minimum as the place where $\loss(\bi{s})$ is 
minimized, and the posterior is maximized globally. 

\subsection{Forward Model}
\begin{figure}
\centering\includegraphics[width=0.7\textwidth]{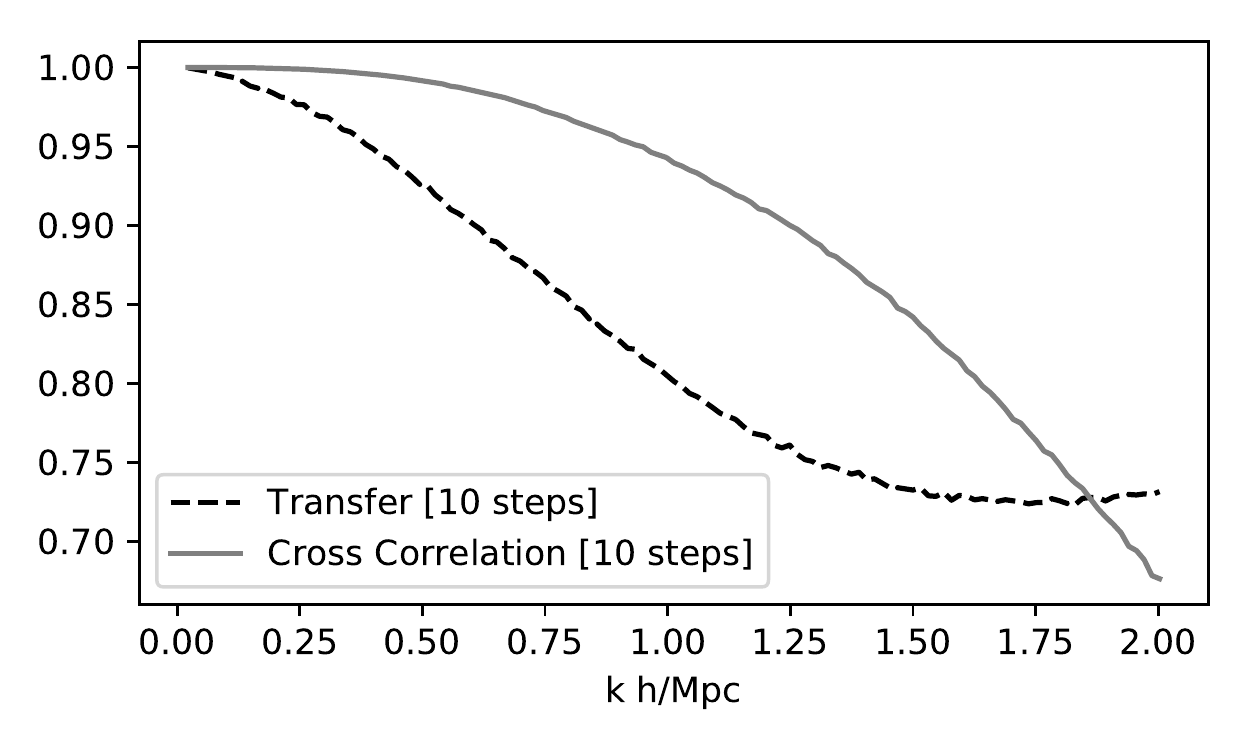}
\caption{Comparing the 10 step forward model against a high resolution simulation.}
\label{fig:100-vs-10}
\end{figure}

The forward model 
$f : \mathrm{LN} \rightarrow \mathrm{NL}$
transforms a linear density field to a non-linear density field. For the purpose of this paper we use a simulation box of L=400 Mpc/h per side, with $N_g^3 = 128^3$ particles. We will 
work with the same number of modes as the number of data points, $128^3$.   
The cosmology is the latest Planck cosmology \cite{Planck15}. We use a 10 step FastPM \citep{FengChuEtAl16} simulation for the nonlinear forward model, with a force mesh boost factor of 1.
We find that at this resolution, using 10 steps is a good compromise between the accuracy and 
the speed/memory consumption, since it is 
able to achieve approximately correct 
nonlinear evolution on the scales of interest here (Figure \ref{fig:100-vs-10}): the cross-correlation 
coefficient against a full N-body simulation is 0.95 at $k=1\hmpc$, which is the 
maximum wave-vector we use in this paper. Employing very few steps also means the 
back propagation algorithm needed to evaluate the gradients is reasonably affordable in terms of memory storage and computing time: we need to store ``only" 10 times the final simulation data. Finally, the transfer function between the 10 step FastPM 
and full N-body is 0.90 at $k=0.6\hmpc$ and 0.8 at $k=1\hmpc$, so the 
loss of power relative to a full simulation is relatively small. \footnote{Using 5 steps gives a very similar accuracy, though most of the experiment in this work is done with 10 steps.} Since we use the same
10 step FastPM to generate both the data and the solutions there is no need 
to correct for this loss of power, but when we compare the results to the 
high resolution simulations we should do so. We note that increasing the number of 
time steps did not improve much the correlation coefficient and transfer function, 
increasing the latter by 2\% at the highest $k$: most of the loss of resolution is 
due to the finite mass and force resolution, which can be improved by going to 
higher mass and force mesh resolutions rather than more time steps. 

\subsection{Noise Model}
\begin{figure}
\centering
\includegraphics[width=0.5\textwidth]{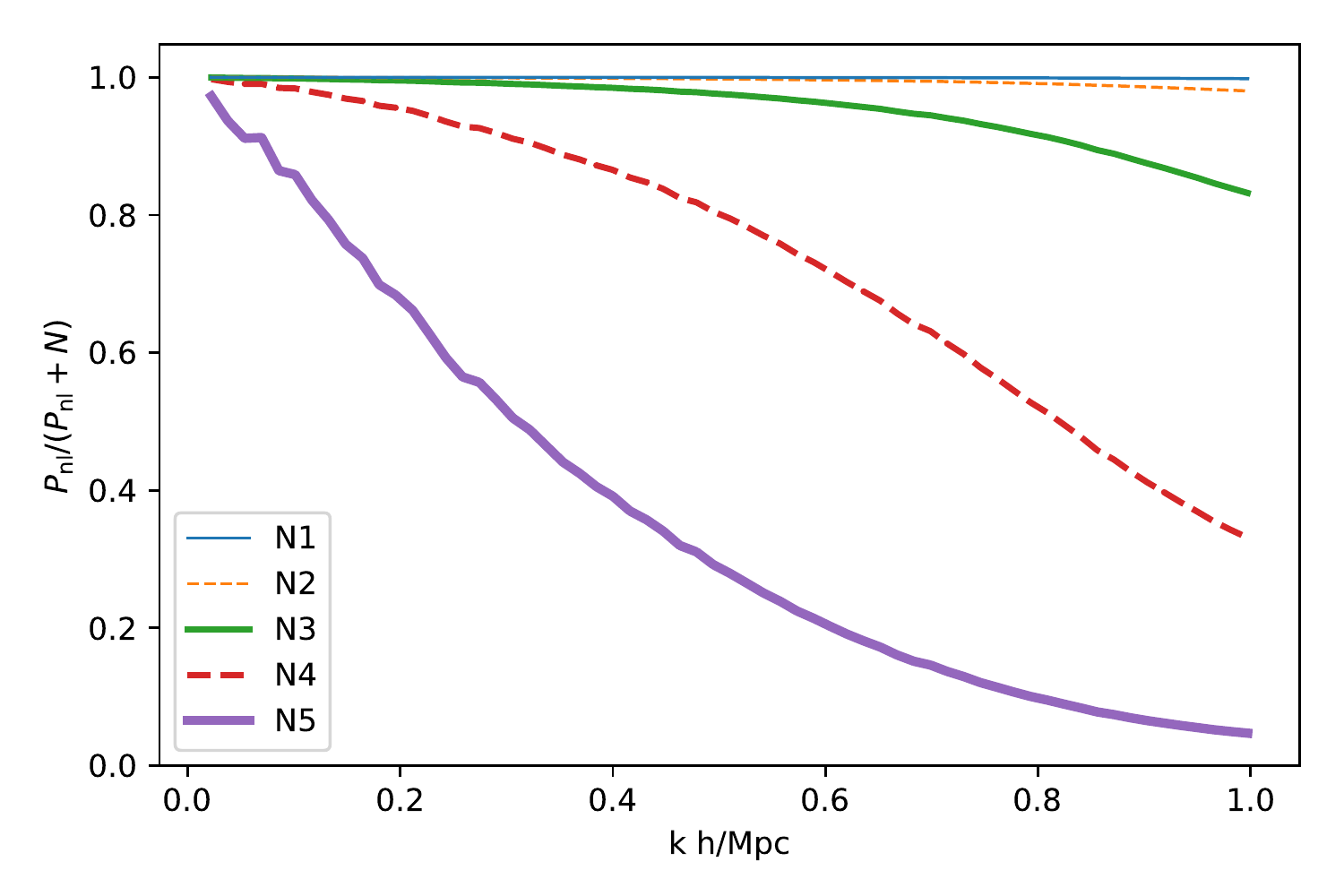}
\caption{Non-linear signal to noise ratio of the simulated datasets. Also see Table \ref{tab:noiselevels}.}
\label{fig:noiselevels}
\end{figure}

The noisy synthetic data ${\bi d}$ are produced with the same forward model $f$,
\begin{equation}
d_i = f_i(\bi{s}) + n_i,
\end{equation}
where $\bi{s}$ is the linear density field at the truth. We generate it as a Gaussian realization with power spectrum $S(k)$. The noise $\bi{n}$ is sampled from a white noise power spectrum with power $N=\sigma_n^2  \left(\frac{L}{N_g}\right)^3$, where $\sigma_n^2$ is the variance of the white noise. We follow the Fourier space sampling scheme of N-GenIC \citep{Springel05} to generate the scale invariant Gaussian realizations from random seeds. 
We produce five datasets for the same truth with noise realizations of different noise power spectrum, varying the power by a factor of ten. The noise levels of 
the five datasets are listed in Table \ref{tab:noiselevels}, and shown in Figure \ref{fig:noiselevels}. The noise levels N1-N3 are unrealistically low when compared to the actual data: typical number 
density of galaxies in a redshift survey rarely exceed $\bar{n} \sim 10^{-2}(\hmpc)^3$ for 
local surveys, and are typically much lower for higher redshift surveys. 
We note, however, mass weighting can reduce the noise below the shot noise level 
$N=\bar{n}^{-1}$, and a survey with a number density of  $\bar{n} \sim 10^{-2}(\hmpc)^3$ can give effective noise three times lower ($N \sim 30(\hmpc)^3$) \cite{HamausSeljakEtAl10}. 

\begin{table}
\centering
\begin{tabular}{cccc}
\hline\hline
Dataset name & Noise Power (Mpc/h$^3$)  \\
\hline
N1   & 0.1    \\
N2   & 1      \\
N3   & 10     \\
N4   & 100  \\
N5   & 1000  \\
\hline\hline
\end{tabular}
\caption{Simulated datasets and their noise level. Also see Figure \ref{fig:noiselevels}. }
\label{tab:noiselevels}
\end{table}
\subsection{Optimizer and starting points}
For any given starting point $\bi{s^0}$ the goal of optimization is to 
descend down the loss function until it finds a local minimum. For 
convex posteriors this is also a global posterior maximum, while for non-convex
functions there is no such guarantee. Optimization routines 
may also get stuck on a saddle point, and in some of the high resolution cases 
our convergence rate has slowed down considerably. 

For this work we use quasi-Newton L-BFGS based optimization \cite{NocedalWright06}
coupled with an adiabatic scheme to go from large scales to small scales (appendix \ref{appa}). The optimization is an iterative procedure, where at every updated position 
$\bi{s^i}$ one chooses a new direction to move based on gradient information 
of the data model with respect to the initial modes $\frac{\partial{\mathcal{L}_d}}{ \partial{\mathbf{s}}}$, plus that of the prior term $\frac{\partial{\mathcal{L}_s}}{ \partial{\mathbf{s}}}$. 
In quasi-Newton's methods one also attempts to approximate the Hessian to 
give a better estimate of the direction of the next step. When such an 
update fails to reduce the loss function (poor Hessian approximation) we revert to the gradient direction. This gives a 
sequence
\begin{equation}
   \left\{\bi{s^{0}}, \dots, \bi{s^{N}}, \dots \right\} \rightarrow \bi{s^{*}},
\end{equation}
and usually truncated at step $N$ under some convergence criteria. Here we truncate when the difference of objective function $y$ between two steps is less than $\epsilon y$, with $\epsilon = 10^{-7}$. There is no guarantee
that this condition is sufficient to converge to a local minimum, and indeed we may have some evidence 
that we have on occasion not completely converged, specially for low noise situations. Typical number of 
iterations is of order several hundred to several thousand. We expect that true second order methods such as 
Steihaug-conjugate gradient \cite{NocedalWright06} will improve the convergence, and we plan to investigate these in the future. 

Non-convex optimization results can be affected by the starting point $\bi{s^{0}}$, 
since the starting point determines which local minimum the optimization will 
descend to. The posterior surface is certainly convex for large scale 
linear modes and (almost certainly) non-convex for small scale modes 
past the shell-crossings, but for any specific level of noise and resolution 
we do not know if the posterior surface is convex or not. For 
this reason we explore convergence as a function of different starting points. 
We consider three choices of starting points: \begin{itemize}
\item zero O: $\bi{s^{0}} \rightarrow 0$. The starting point is chosen to be close to 0, where the prior term is already minimized. We do not start at exactly 
$0$ because this is a local stationary point with a vanishing gradient.
\item truth T: $\bi{s^{0}} = \bi{s}$. Starting from the linear density field of the truth. Because of the prior, the truth is not a minimum of the 
loss function except for zero noise case. This starting point is of course unrealistic for real data 
analysis, but 
it can help quantify the nature of the posterior surface. While it may or 
may not find the global minimum, it is likely to be close to the minimum of 
the data likelihood term, since it starts at the truth. 
\item random R: $\bi{s^{0}}$ is a different seed realization with the same power spectrum as the one used to sample the truth. We consider two random starting points, R1 and R2, with different random seeds.
\end{itemize}

\section{Numerical results: N-body simulations}
\label{sec2}
\subsection{Reconstruction results: effects of noise and starting point}

\begin{figure}
\centering
\includegraphics[width=0.47\textwidth]{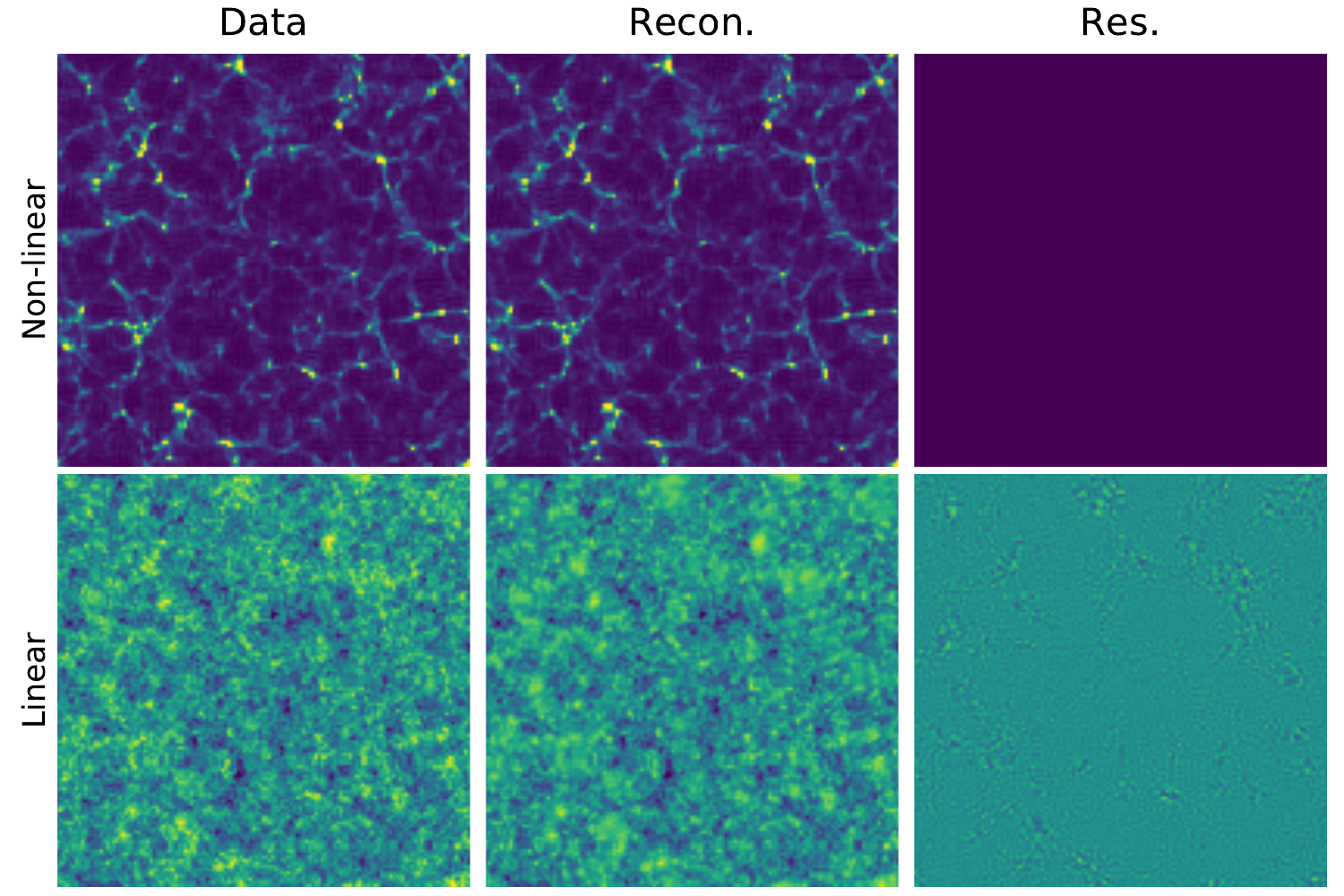}
\includegraphics[width=0.47\textwidth]{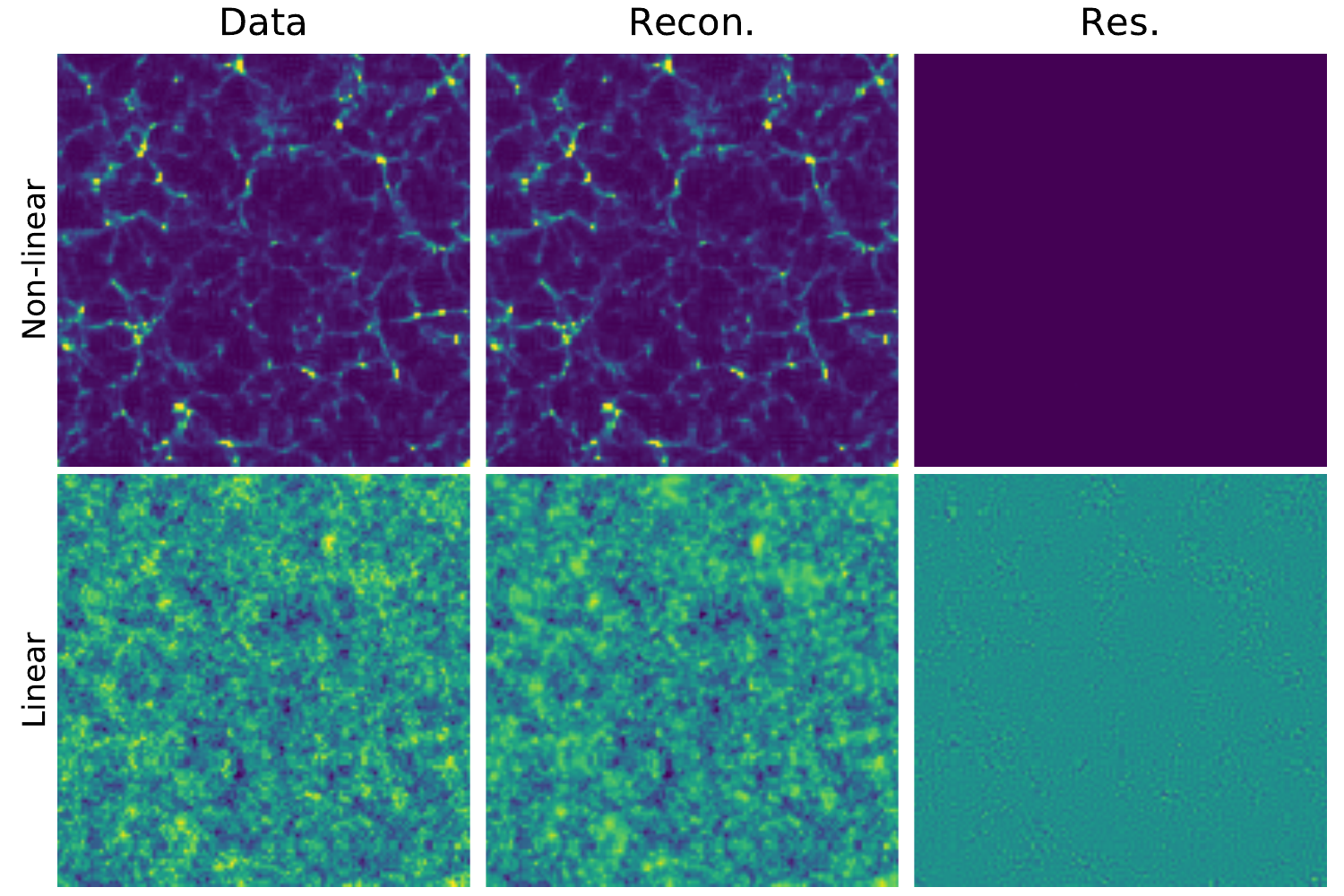}
\caption{Visual impression of the low noise case (N1). The left panel is starting from zero, the right panel is starting from truth. Color scales fixed for each horizontal row. Both give a perfect reconstruction of the data, and, although not perfect since there is 
some non-zero residual,
we cannot visually see the difference between the two reconstructed initial fields.}
\label{fig:visualN1}
\end{figure}

\begin{figure}
\centering
\includegraphics[width=0.47\textwidth]{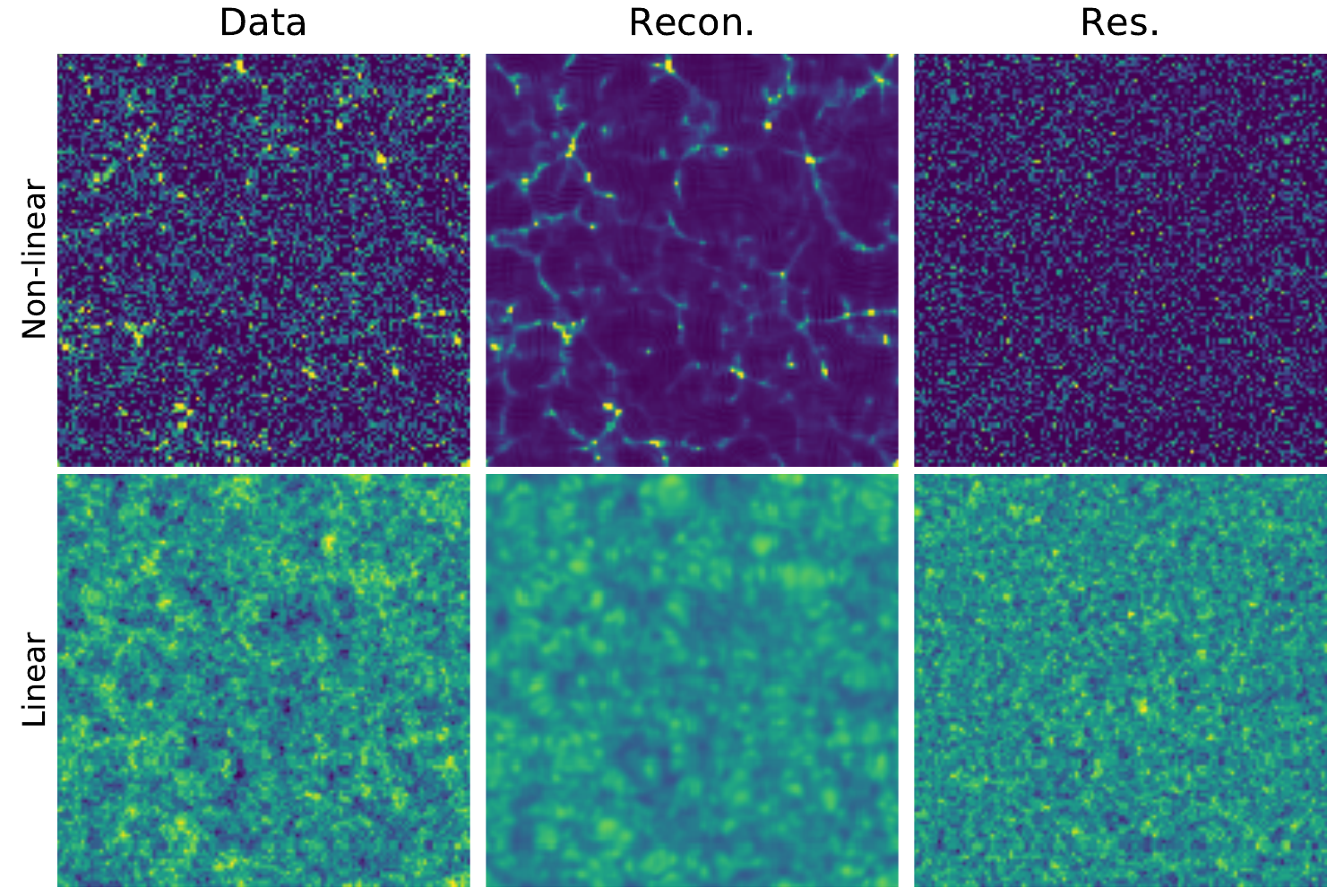}
\includegraphics[width=0.47\textwidth]{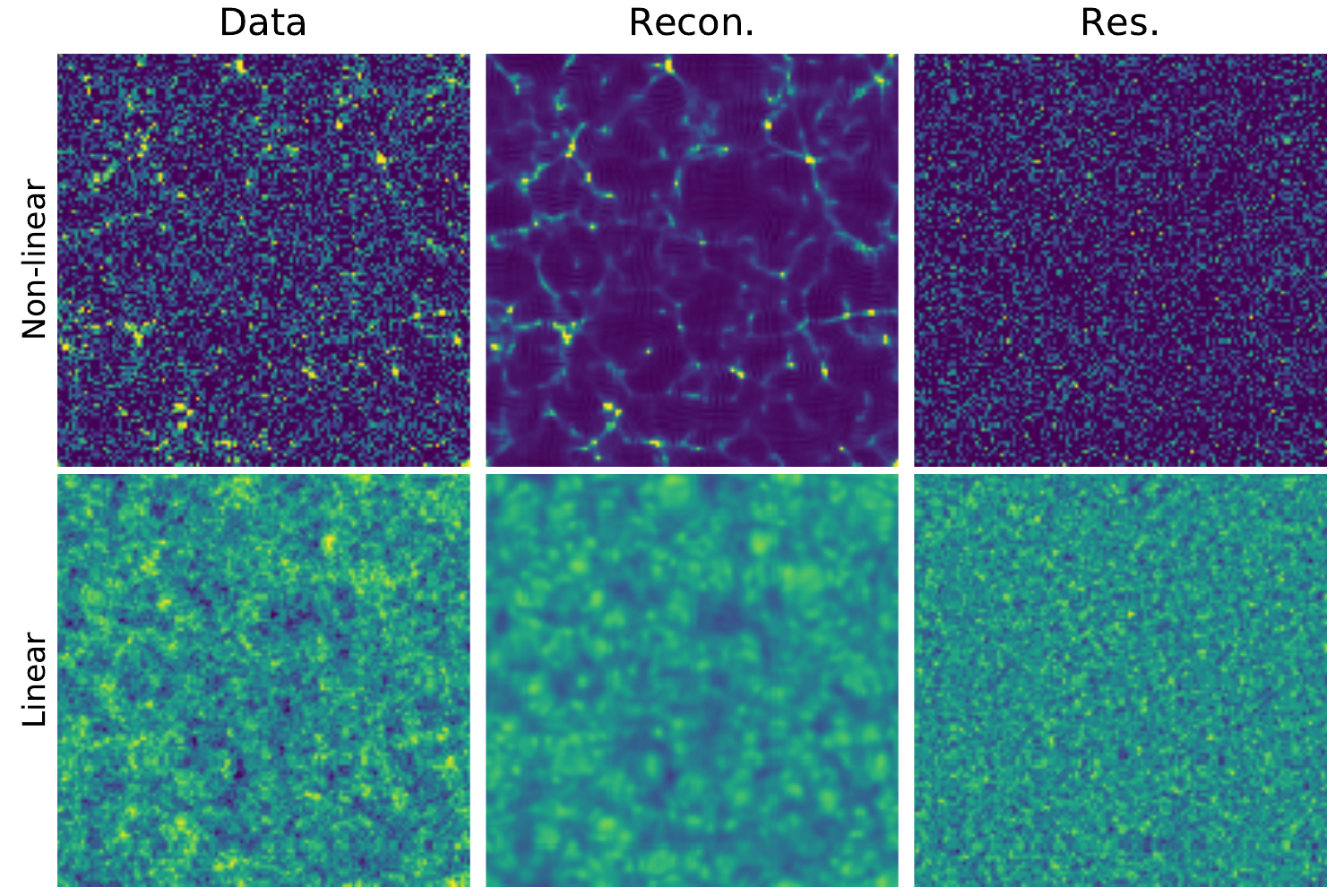}
\caption{Visual impression of the high noise case (N5). The left panel is starting from zero, the right panel is starting from truth. Color scales fixed for each horizontal row. In this case the difference between the two linear reconstructions is more evident due to a larger noise component, and residuals are larger for starting from zero case. In both cases the linear reconstruction is more heavily smoothed than in the low noise case, since reconstruction sends noise dominated modes to zero.}
\label{fig:visualN5}
\end{figure}

We begin exploration of our results by showing a 
visual impression of the reconstruction procedure after the convergence 
criterion has been met. We show the reconstruction of the low noise (N1) in figure \ref{fig:visualN1} 
and of the high noise (N5) datasets in figure \ref{fig:visualN5}, with starting points at zero and truth 
for both cases. 
The plots show a 10$\hmpc$ slab in radial direction with the full $(400\hmpc)^2$ extent in transverse 
directions. 

In the low noise case (figure \ref{fig:visualN1}) 
we find an essentially perfect reconstruction of the data, which can be seen 
from the fact that the residuals of the data are nearly zero for both starting points. 
This shows that our reconstruction code can converge to a near perfect result.
In the high noise case (figure \ref{fig:visualN5}) the two starting points also lead to a very similar result both in terms of 
the final data reconstruction and in terms of the initial linear field reconstruction. The latter is 
more heavily smoothed than before, a consequence of noise having high frequency power. However, the data
reconstruction is visually quite good, as can be seen by comparing to the no noise 
truth as the data column in figure \ref{fig:visualN1}. 
The residuals of the linear field seem to depend on the starting point: they appear to have a more correlated structure when starting at zero. 
These results suggest a presence of multiple minima in the loss function 
(i.e. multiple maxima in the posterior), but these appear to be high quality minima, in the sense that 
the final reconstructions are similar, yet the linear solutions are distinct. 
We quantify this further below.

\begin{figure}
\centering\includegraphics[width=0.7\textwidth]{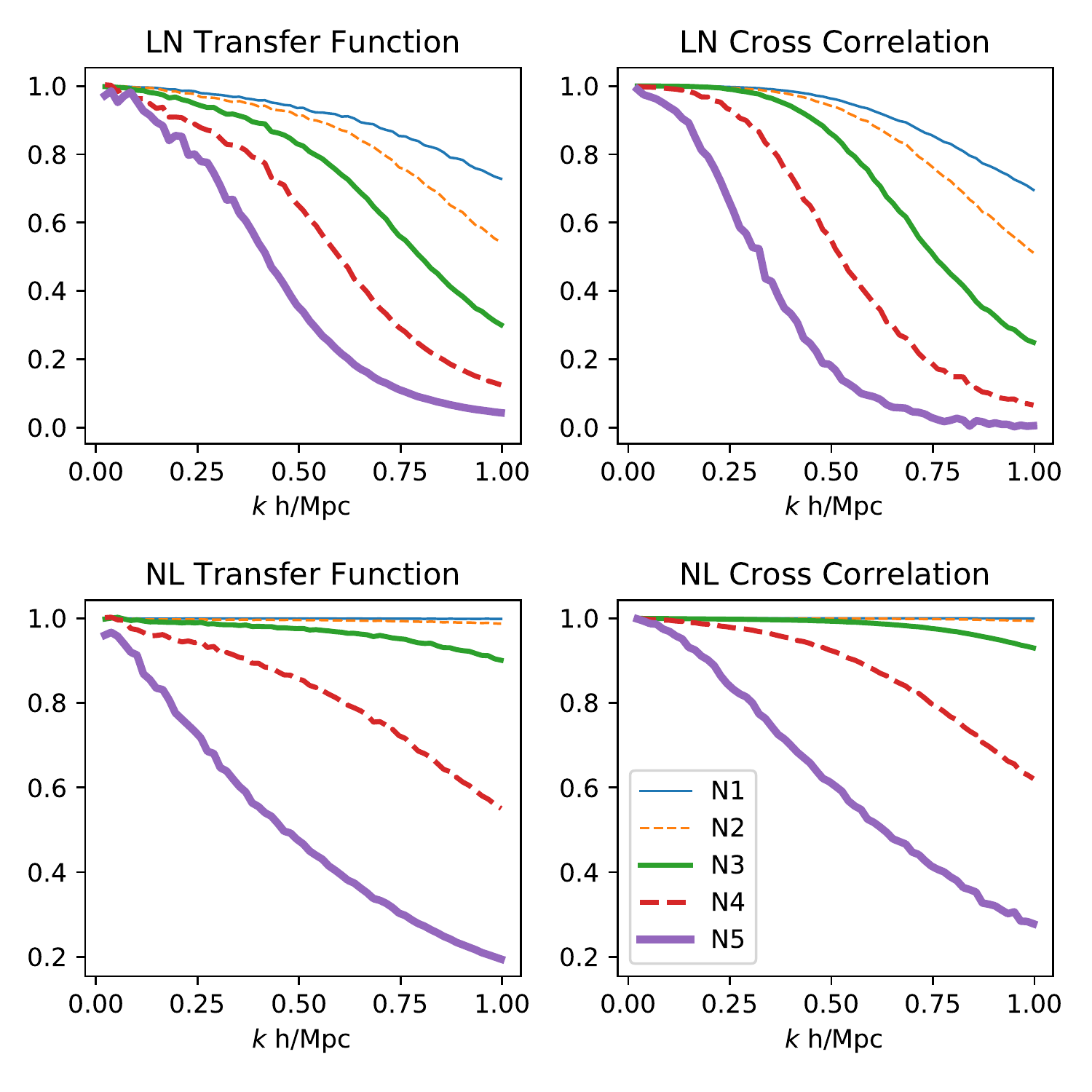}
\caption{Reconstruction results for the transfer function and correlation coefficient, starting at random (R, with the correct power), for varying noise levels.
The linear reconstruction (LN) is compared to the true input, while the nonlinear reconstruction (NL) is compared against noisy data. }
\label{fig:noise-conv-random1}
\end{figure}

\begin{figure}
\centering\includegraphics[width=0.7\textwidth]{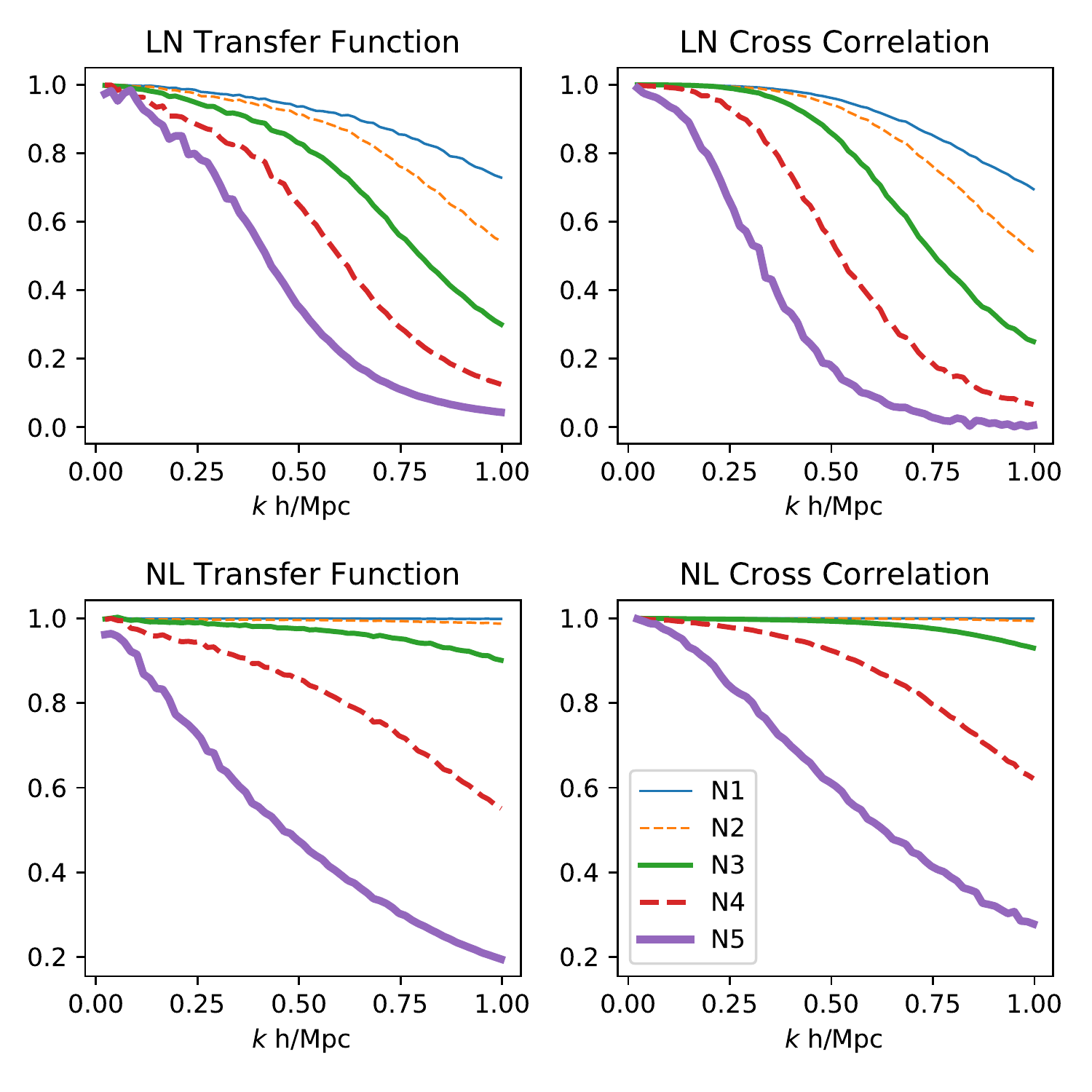}
\caption{Reconstruction results for the transfer function and correlation coefficient, starting at (near) zero (O), for varying noise levels. 
The linear reconstruction (LN) is compared to the true input, while the nonlinear reconstruction (NL) is compared against noisy data.}
\label{fig:noise-conv-zero}
\end{figure}

\begin{figure}
\centering\includegraphics[width=0.7\textwidth]{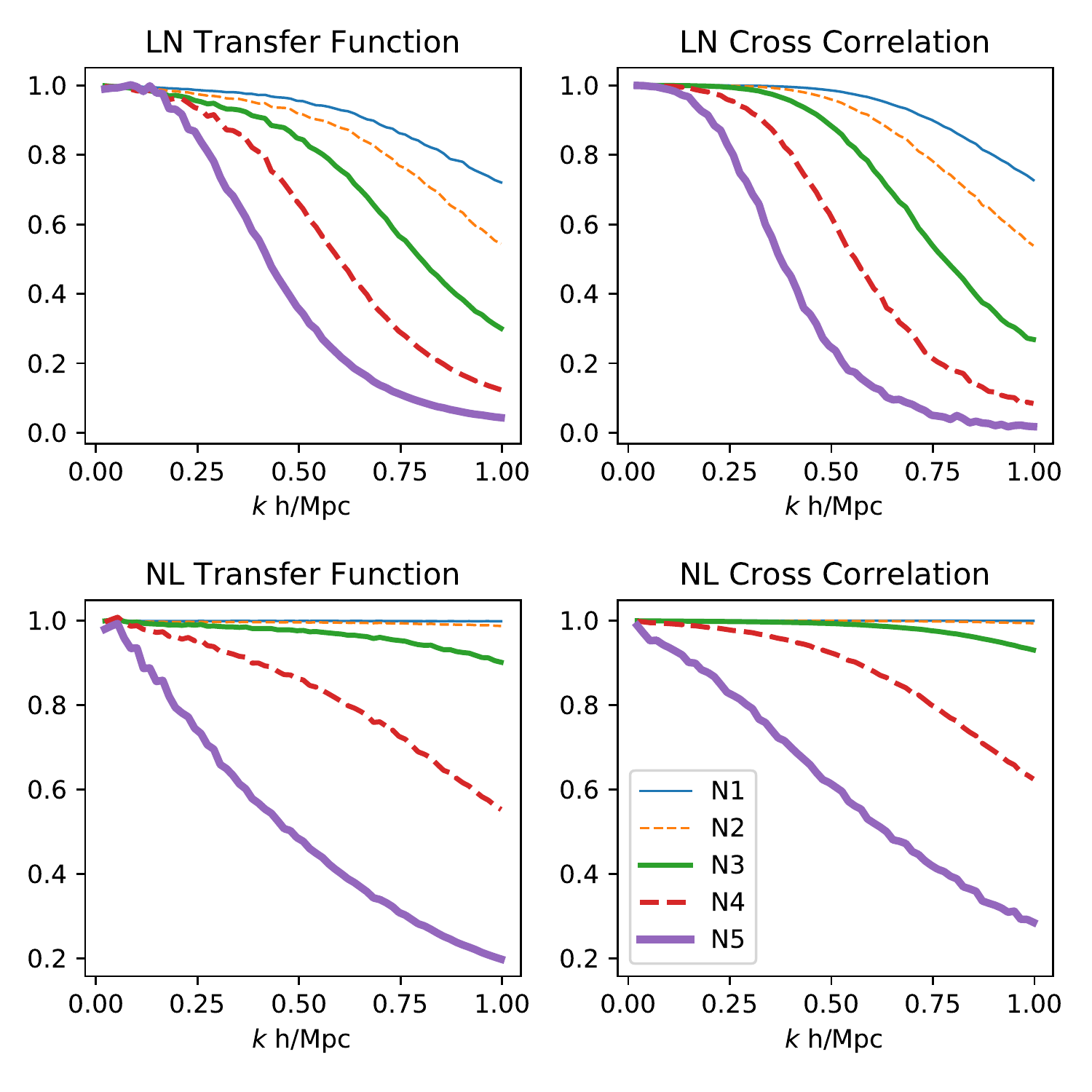}
\caption{Reconstruction results for the transfer function and correlation coefficient, starting at truth (T), for varying noise levels.
The linear reconstruction (LN) is compared to the true input, while the nonlinear reconstruction (NL) is compared against noisy data. }

\label{fig:noise-conv-truth}
\end{figure}

We have seen that the level of noise affects the initial reconstruction regardless of the starting point. In figure \ref{fig:noise-conv-random1} 
we show the results of the reconstruction applied to the datasets with varying 
noise, for the random starting point. We introduce two statistics to quantify the results. 
The transfer function (TF) is defined as the square root of the ratio of the power spectra, $\sqrt{\frac{P_\mathrm{recon}}{P_\mathrm{truth}}}$ while the cross correlation (CC) coefficient is defined as 
the ratio of the cross-power between reconstruction and truth 
to the square root of the product of the truth and reconstruction power spectra, 
$r_{cc}=P_\mathrm{recon,truth}/\sqrt{P_\mathrm{recon}P_\mathrm{truth}}$.
Both of these statistics are affected by the noise: when the noise is higher the 
reconstruction becomes worse, and the solution is driven to a lower amplitude
at high $k$. 
As we reduce the noise the reconstruction of nonlinear (NL) data 
becomes near perfect, first on large scales and then increasingly so on small scales. 
The main reason for this scale dependence is the shape of the power spectrum of LSS
relative to white noise: LSS has a red power spectrum, i.e. a lot 
of power on large scales (low $k$), and power decreases towards small scales. At some scale the LSS power drops 
below the noise power, the data become noise dominated and we can no longer reconstruct the 
initial modes. This picture is modified somewhat because of the nonlinear transformation between the 
linear modes and the nonlinear data model, so that the scale at which noise equals signal can no longer 
be simply determined from the noise and signal power spectrum comparison. 

Figure \ref{fig:noise-conv-random1} also shows that there is considerable difference between the reconstructed
linear (LN) and nonlinear (NL) fields
. While the NL reconstruction is essentially perfect for the low noise cases, the linear reconstruction is 
considerably worse. The physical origin of this is the transfer of power from the large scales to the small scales
during the nonlinear evolution. This is exemplified by the halo collapse model, where a given 
region of mass that is initially nearly homogeneous collapses due to gravity into a very nonlinear object, effectively compressing the structures to scales smaller 
than the ones we look at here. Thus one reason for not being able to reconstruct 
fully the linear modes at high $k$ is that those have been mapped into small 
scales, below the smoothing scale of our data ($3\hmpc$ cells with CIC mass assignment of each 
particle to the nearest 8 cells for our 
experiments). 

Transfer of power from large to small scales is however not the only process occurring during the halo collapse: 
there are also considerable shell crossings of particles that can effectively erase the memory of initial conditions, specially  
if only density observations are used. 
Whether the shell crossings matter or not also depends on the scales we are interested 
in. If we have shell crossings and information is effectively destroyed this can 
be an alternative explanation why we are unable to reconstruct fully the linear 
modes on small scales. We discuss this further below. 

We note in passing that 
with a full phase space information (density and velocity) there would be, in principle
at least, 
no such erasure of information (on scales larger than the coarse-graining scale of observations), 
since the system is deterministic and reversible. In cosmology we however very rarely have 
information on both density and velocity, so we will not explore this scenario here. 
Note that it is the special initial conditions in cosmology, where the initial velocity 
and the initial displacement are equal (in dimensionless units) that allows us to do a 
reconstruction from just half of the phase-space data, without ever using velocities.

\begin{figure}
\centering\includegraphics[width=0.7\textwidth]{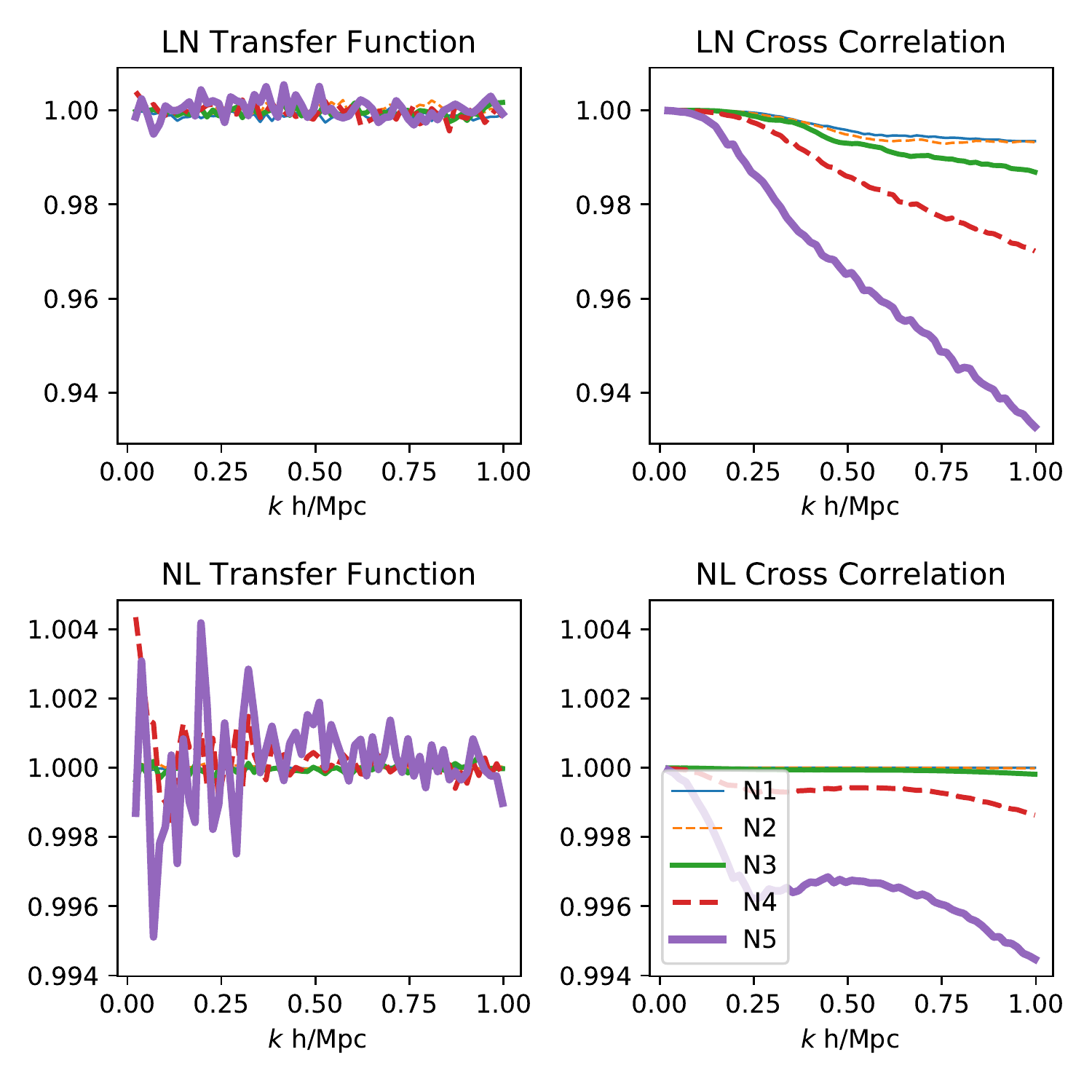}
\caption{Comparing reconstruction results of
starting points zero (O) and random (R), for varying noise levels. The distance between the two solutions are measured as the relative transfer function (R/O) and cross correlation coefficient.}
\label{fig:random1-vs-zero}
\end{figure}

\begin{figure}
\centering\includegraphics[width=0.7\textwidth]{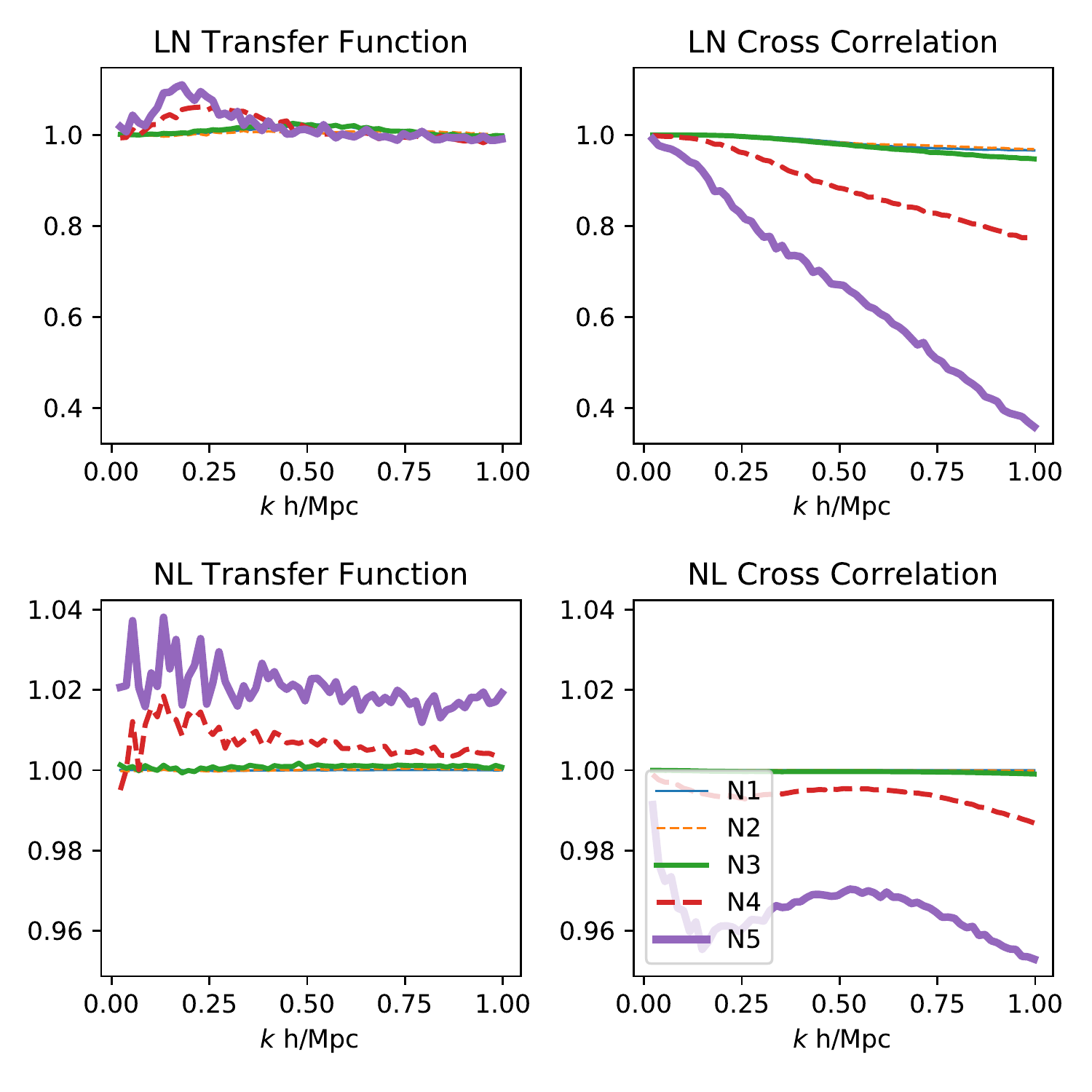}
\caption{Comparing reconstruction results of starting points at truth (T) and random (R), for varying noise levels. The distance between the two solutions are measured as the relative transfer function (T/R) and cross correlation coefficient.}
\label{fig:random1-vs-truth}
\end{figure}

In figures \ref{fig:noise-conv-zero} and \ref{fig:noise-conv-truth} we show 
the corresponding solutions 
for starting from zero and from truth. They are qualitatively similar to the starting from random case. 
To understand the differences between the different starting points we compare them 
to each other. Random versus zero is shown in 
figure \ref{fig:random1-vs-zero}. In the latter 
case there is almost no difference between the 
solutions in terms of the transfer functions, both in the linear (LN) and in the nonlinear field (NL): this is saying that all solutions are equally 
good on the data reconstruction, giving the same transfer function. We find the same result when comparing two different random 
starting point solutions (shown further below). This is good for the 
reconstruction method: different realistic starting points all lead to the solution with the same power, so we  are able to correct the missing power by calibrating the procedure on simulations. The cross-correlation coefficients differ 
somewhat in the high noise regime, suggesting the solutions are not identical. This could be either due to insufficient convergence 
or existence of multiple minima, but the differences are small and in any case the 
solutions appear to be equivalent in terms of their quality when compared to truth.

When comparing the truth to the random (or zero) starting points, shown in 
figure \ref{fig:random1-vs-truth}, we see that in most cases the transfer 
functions are also very close to each other, with the exception of the highest two 
noise cases where 
truth has higher TF than random starting point, by up to 10\% in linear case. Surprisingly, this difference is 
at relatively low $k$ values, caused by slow convergence of large scale modes in the 
high noise regime when starting at random or zero. This could be a numerical issue and 
deserves further exploration. 


The cross-correlation coefficient is close to one for low 
noise, but significantly away from one for high noise. For high noise the results again suggest that 
the solutions are not the same and the posterior surface is thus non-convex, converging to a different 
local minimum. Note however that in terms of quality of solution starting from truth does not lead to 
a much better solution than starting from either zero or random, as seen from figures 
\ref{fig:noise-conv-random1}-\ref{fig:noise-conv-truth}. 
So while the optimization solutions may have only found a local minimum, 
they are of comparable quality and anyone of these solutions can give a 
good power spectrum estimator, as long as they can be calibrated in terms of transfer
functions. 

\subsection{The loss function}

\begin{figure}
\centering\includegraphics[width=0.32\textwidth]{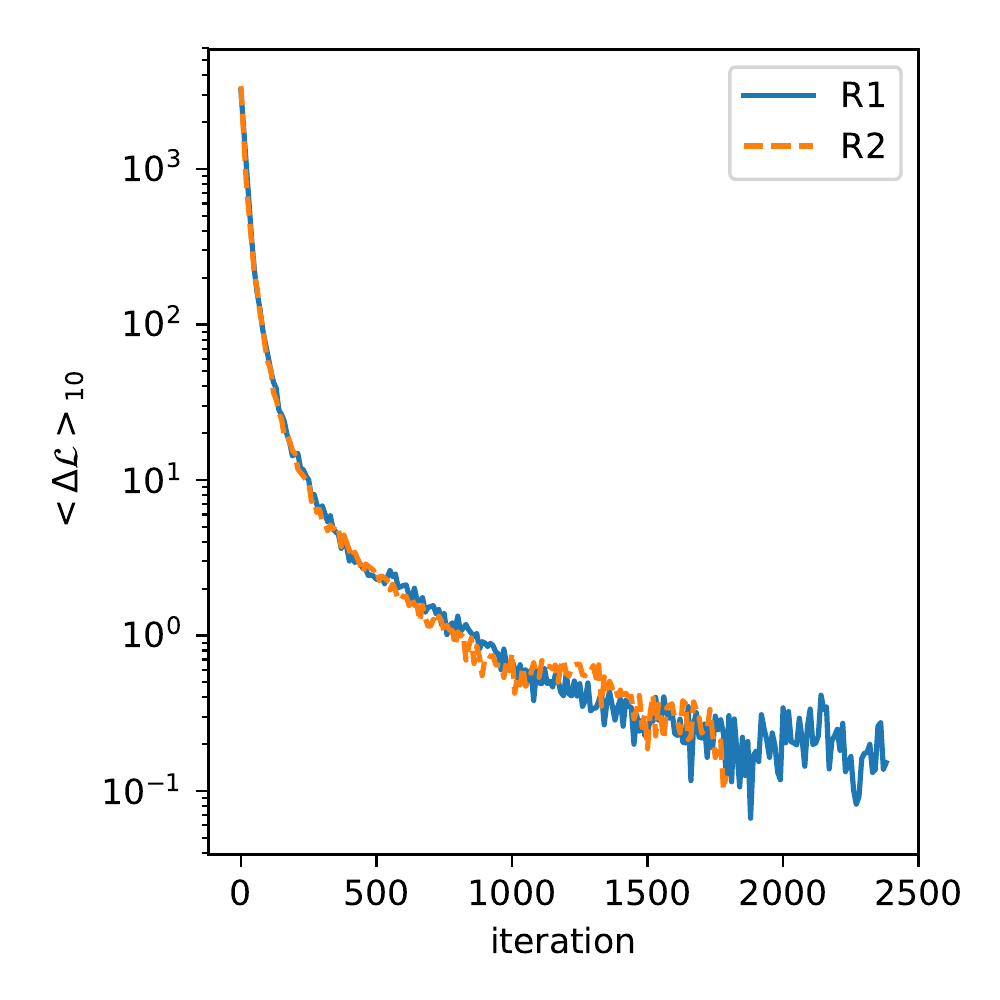}
\includegraphics[width=0.32\textwidth]{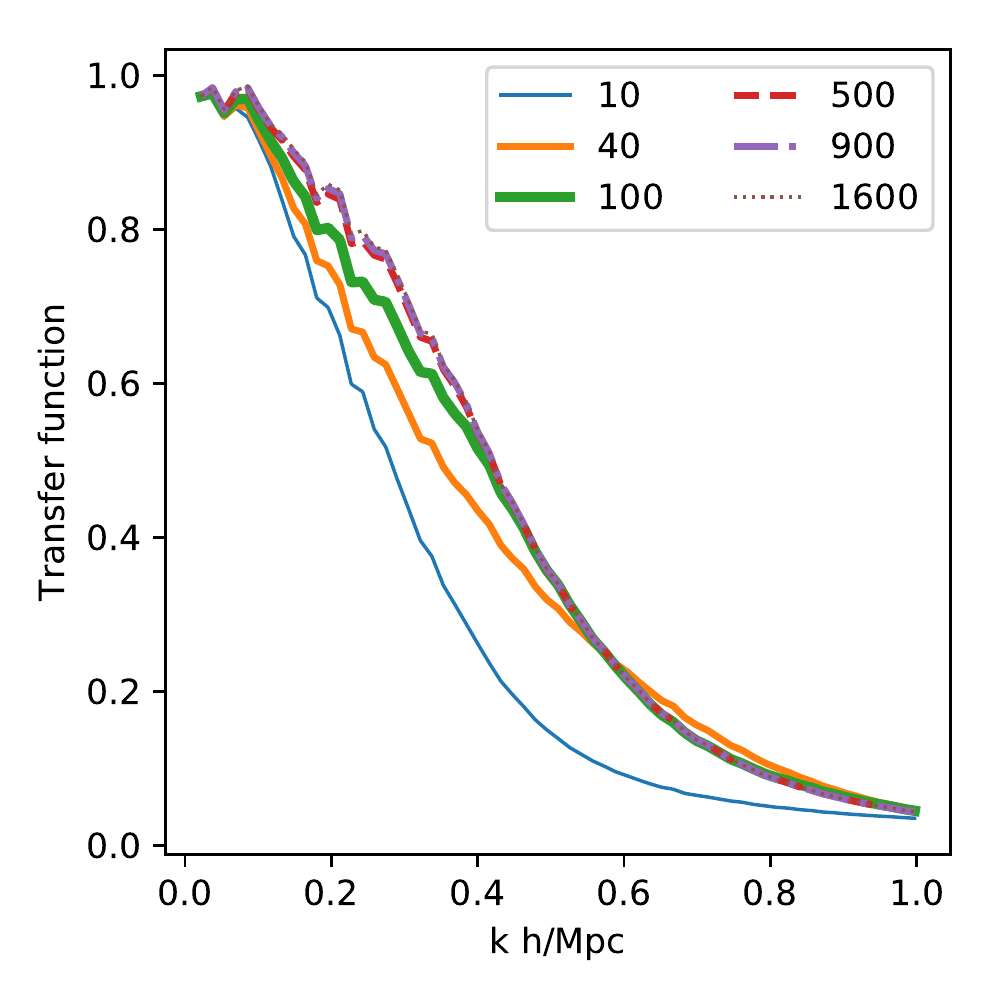}
\includegraphics[width=0.32\textwidth]{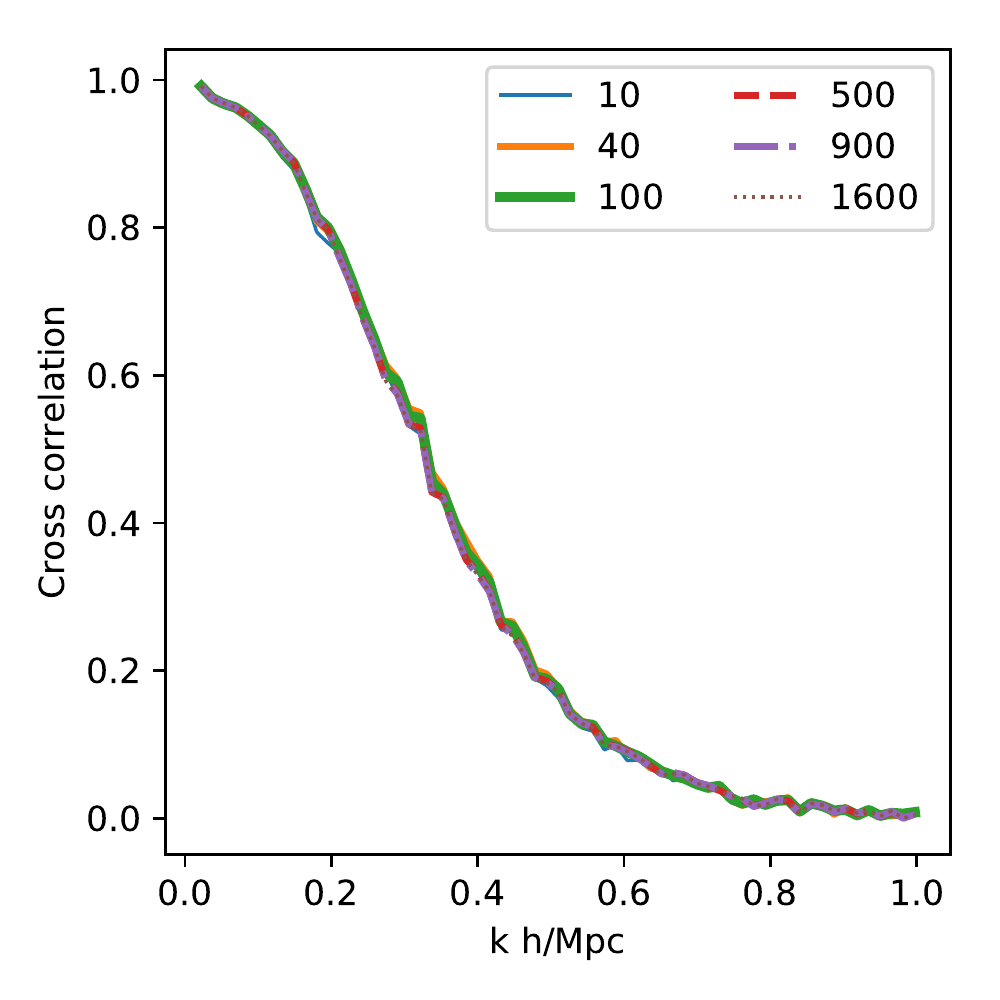}
\caption{Convergence of the loss function and the transfer function as a function of 
number of iterations for N5 noise. Left panel: the average relative descent rate of the loss function of two runs. We see that the loss function have essentially stopped decreasing after 1500 iterations, with a rate around $10^{-7}$. 
Middle panel: Transfer function has converged after 500 iterations. Right panel: the cross correlation coefficient has converged at the very beginning. Note that we only showed the iterations of the final objective function. See the appendix on the adiabatic series of objective functions.}
\label{fig:loss-transfer-iterations}
\end{figure}

In figure \ref{fig:loss-transfer-iterations} we show the convergence of the loss function against the number of iterations. After 500 iterations, the rate of descent reduces to $10^{-6}$, which amounts to about $\Delta \mathcal{L} \sim 1$ per iteration. After 1500 iterations, $\Delta \mathcal{L} \sim 0.1$ per iterations, and successive iterations can no longer introduce substantial descent of the loss function.

In the right panels figure \ref{fig:loss-transfer-iterations}, we also show the evolution of the transfer function and cross correlation coefficient as a function of number of iterations. We observe that the the transfer function shows behavior expected from the objective function: after 500 iterations the transfer function converges at the most important scales. Interestingly though, the cross-correlation coefficient stopped improving even though the objective function is rapidly changing. We note that the optimizer is not simply applying a transfer function correction while keeping the phases fixed, instead the phases 
are also changing a lot, which we verified by cross-correlating 10th and 1000th iteration. 

\begin{figure}
\centering\includegraphics[width=\textwidth]{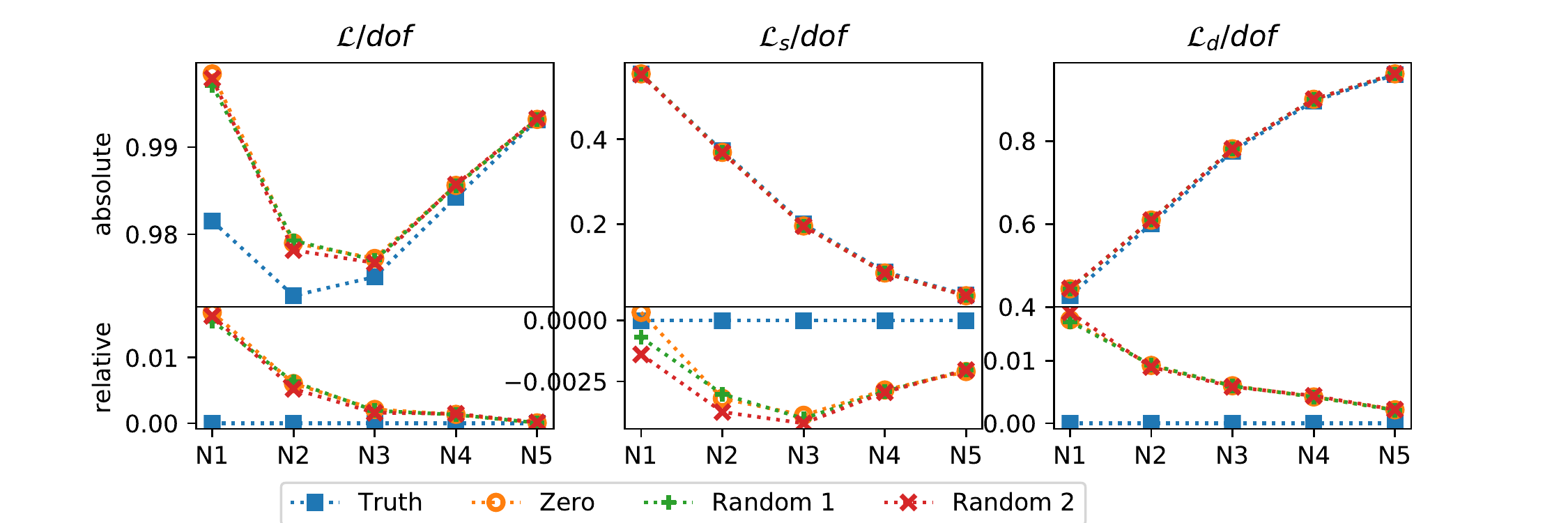}
\caption{Comparing the loss function $\loss$ per degree of freedom (left panel), -2 log prior per degree of freedom (middle panel) and -2 log likelihood per degree of freedom (right panel), for different starting points and different noise levels. $\loss_s$ refers the first term of \ref{eq:obj}, and $\loss_d$ refers to the second term. The lower panels are relative to the case with starting point Truth.}
\label{fig:loss}
\end{figure}

To understand better the quality of solutions we compare in figure \ref{fig:loss} the
loss function $\loss$ divided by the 
number of modes ($128^3$) for different 
starting points. We also show separately the contributions from the prior and the log-likelihood (residual). 
Starting at the truth gives us a lower total loss function in all cases, although the difference vanishes 
for the highest 
noise. This suggests that starting at the truth converges to the lowest minimum (which may or may not be 
the global minimum) and that other starting points lead to a local minimum only. However, as we discussed above, 
starting from the truth is not possible in practice, and all the other starting points lead to the same value of 
loss function, so even if there are multiple minima in the loss function the quality of the attainable 
solutions is always 
the same. This is also true if we start at a different random, and we show that the two random
starting points lead to the nearly identical loss function. We emphasize again that it is difficult to 
separate the lack of convergence from the existence of local minima. We terminate the runs after per-step improvement on the objective function is below $10^{-1}$ for 6 iterations ($\sim 10^{-7}$ per degree of freedom). We find that the low noise case requires more evaluations than the high noise case. The N1 case used as high as 5,000 forward model calls (5,000 FastPM simulations). It still is possible that with a larger number of steps the loss function would further improve. We think this is however unlikely, and the more plausible
interpretation of the results is that the loss surface is mildly 
non-convex, and the global minimum cannot 
be reached using gradient optimization from a random or zero starting point. 
In this paper we do not explore methods that go beyond the gradient based 
optimization, such as annealing methods: it is dubious that such methods 
can be very effective in a very high number of dimensions used here. In any case, 
even if the solutions differ in terms of the loss function, they do not differ much 
in terms of the transfer function and correlation coefficients, which are ultimately
more important than the loss function itself. We have also seen in figure 
\ref{fig:loss-transfer-iterations} that in terms of the transfer function one converges after 500 iterations (and 
in terms of the correlation coefficient a lot less than that), so one does not need to 
have 5,000 iterations for any practical purposes. Other noise levels are qualitatively similar. 

As previous figures show the ``global" minimum (more precisely, our best minimum, but for 
simplicity we will continue to call it global minimum)
is not considerably better from the other minima in terms of the transfer 
function and correlation coefficient. This is further seen in figure \ref{fig:loss}, where the 
global minimum is at most 2\% lower than the other minima in terms of the loss function, 
and hardly any different for high noise levels. The
second panel in figure \ref{fig:loss} shows the prior term, which is actually higher for the global 
minimum due to the global minimum having a higher transfer function (figure \ref{fig:random1-vs-truth}). 
Third panel shows the quality of the fit to the data (residual): here the global minimum is better than the other 
starting points, although the differences are small and vanish for high noise. 

\subsection{Gaussianity of the solution}

\begin{figure}
\centering
\includegraphics[width=0.7\textwidth]{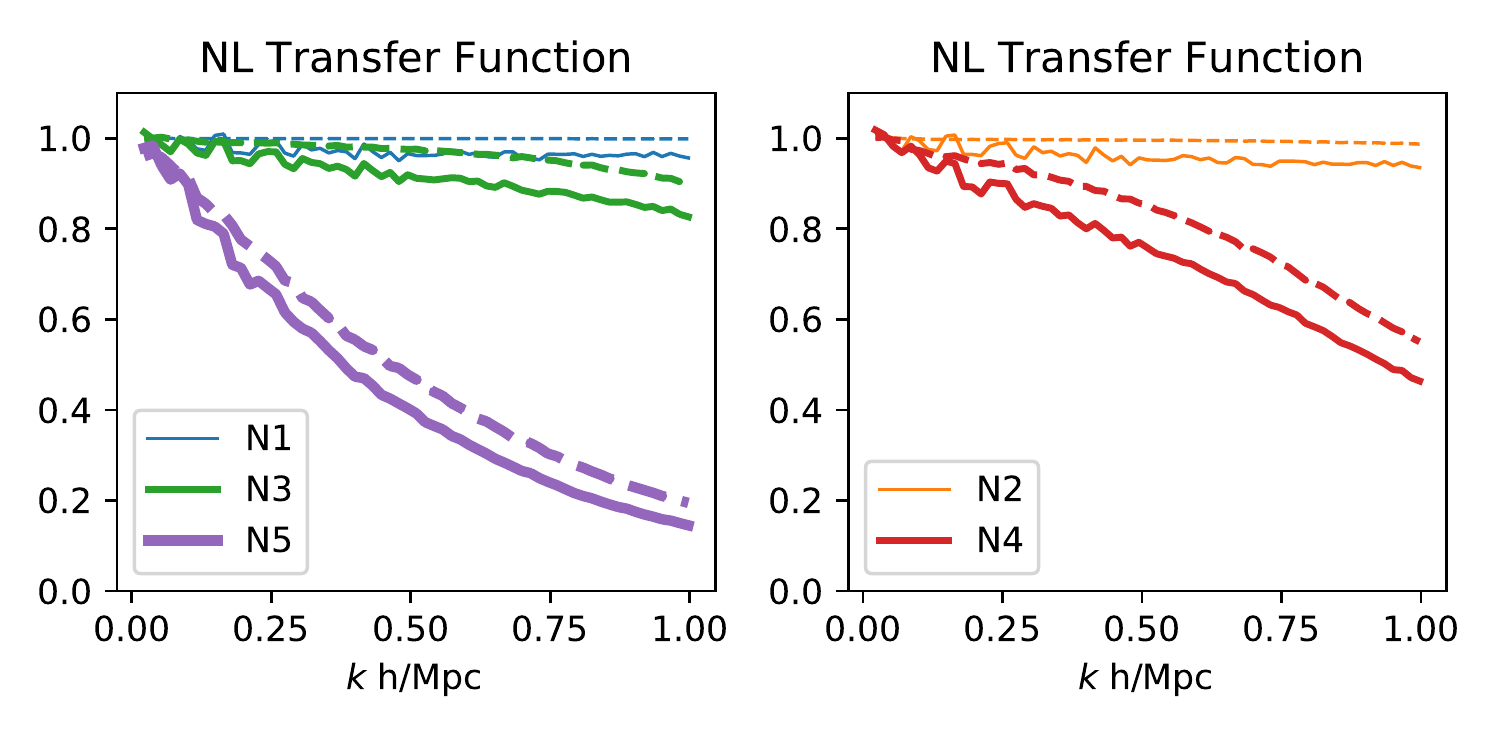}
\centering
\includegraphics[width=0.7\textwidth]{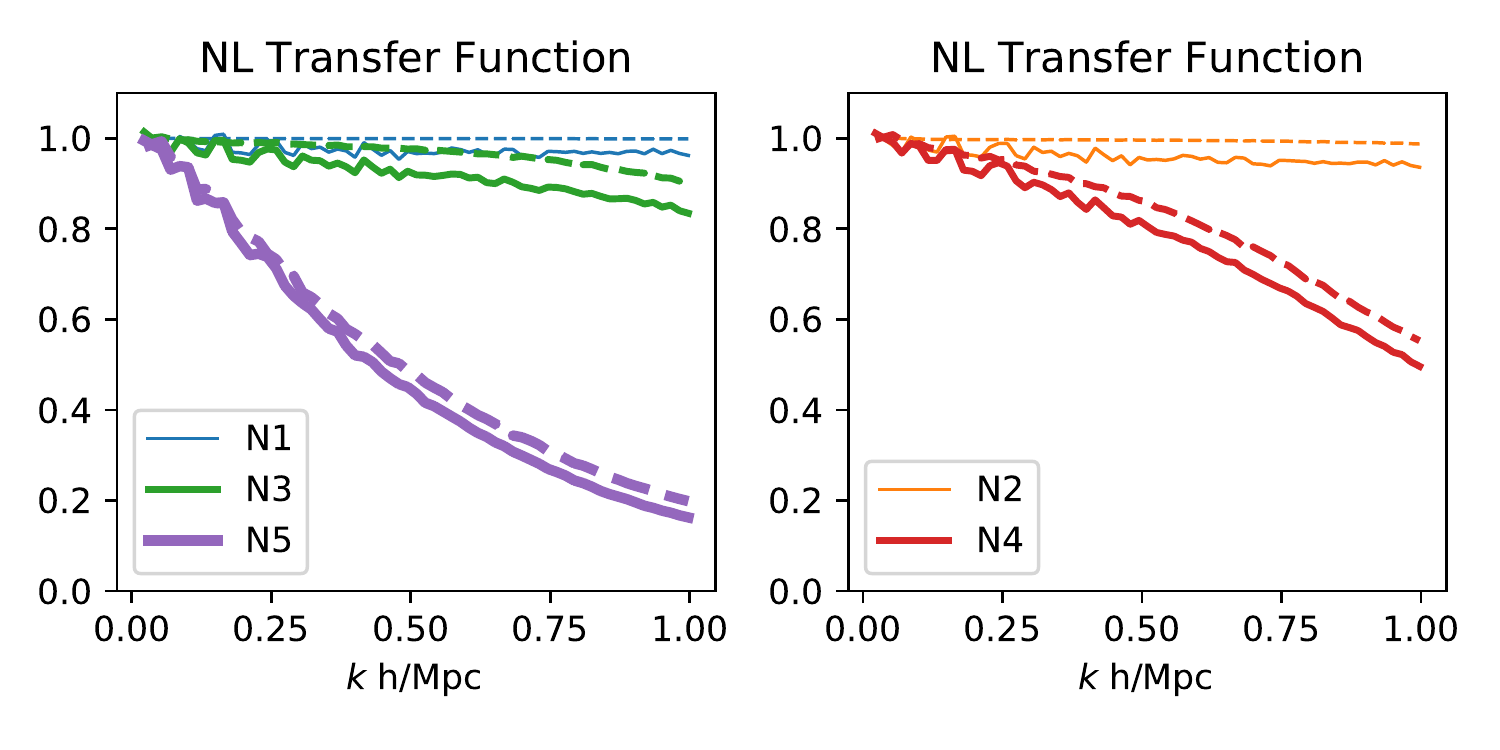}
\caption{Comparing the optimal reconstruction (dashed) against randomizing phases (solid), for starting point at random (R, top) and starting point at truth (T, bottom). We see that although the linear field of the randomized solution has the same amplitude, the power of the non-linear field decreases, for all noise-levels. This indicates the solution is special. }
\label{fig:meddle-random1}
\end{figure}

In figure \ref{fig:meddle-random1} we compare the transfer functions of a reconstructed solution forward modeled, against the same solution where the phases have been randomized. For gaussian initial conditions, randomizing the phases shall not affect the amplitude of the non-linear power spectrum\citep{1991MNRAS.253..295L}. In our case however, phase randomization reduces the power, suggesting these solutions 
are somewhat special, and likely not completely gaussianized. The effect is of 
order a few percent and increases with higher noise levels. As expected the effect 
is more pronounced for solutions starting at random or zero than for those starting 
at truth, but even the latter shows the effect.

\begin{figure}
\centering\includegraphics[width=0.7\textwidth]{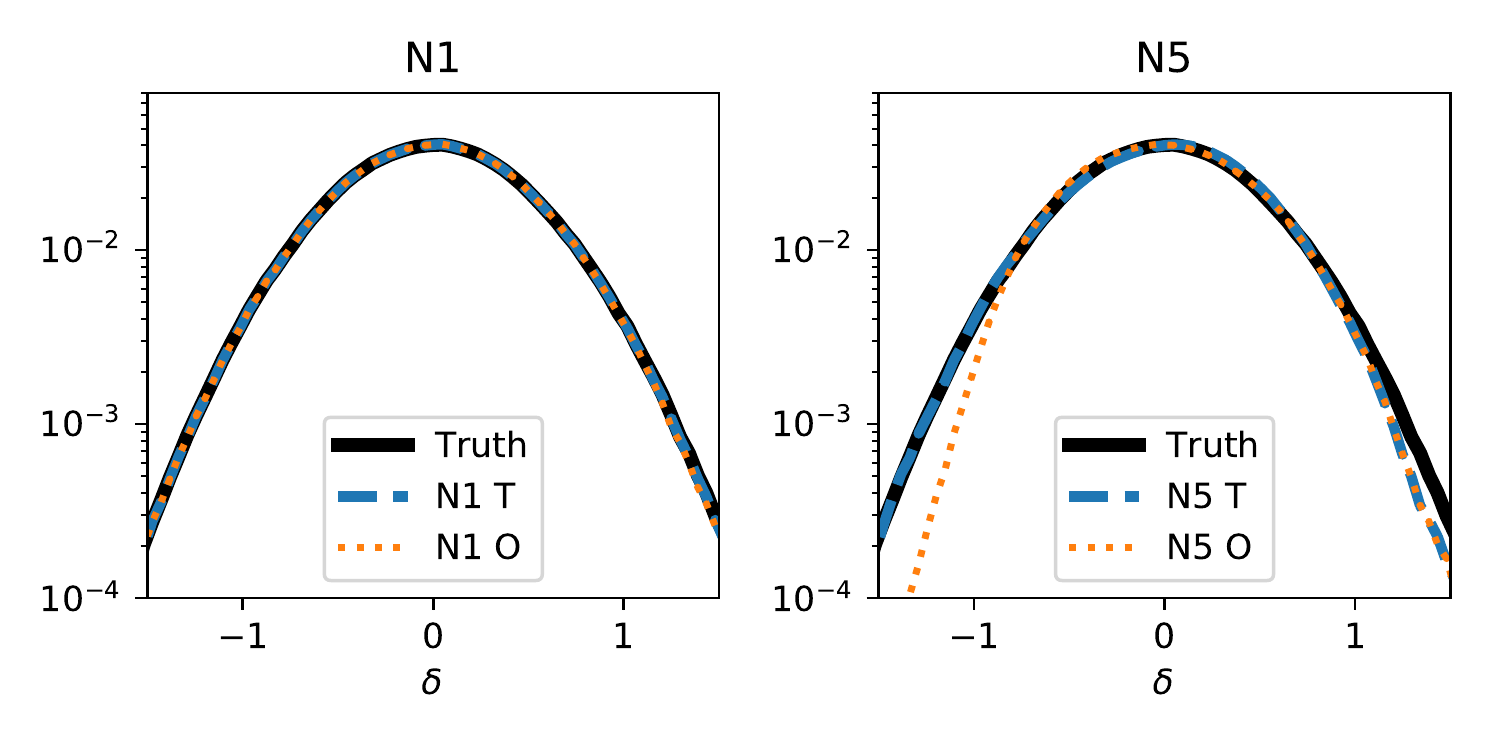}
\caption{Comparing the PDF of the reconstruction at two noise levels, for starting points at zero and truth. As a reference we also show the PDF of the true linear density field, which is Gaussian. We see that for high noise the reconstruction is missing high density peaks, and the solution starting from zero is missing low density peaks as well. The linear fields are smoothed by a 8 Mpc/h Gaussian filter.}
\label{fig:pdf}
\end{figure}

Figure \ref{fig:pdf} explores a related effect, the one point distribution 
function (PDF) of the linear reconstruction at the pixel level. For low 
noise this agrees well with the true initial conditions for all solutions. 
For high noise this is also true for the solution starting at truth, but only 
for low or negative density. The 
solution of starting point zero or random is however significantly non-gaussian
in terms of its PDF and the same is true for solutions of starting point truth at the 
over-densities above one: reconstructed solutions do not reach as high over-densities 
as the truth and this is true even if we start from the truth. 
We conclude that the reconstructions contain some residual level of 
non-gaussianity in the case of high noise. This is not surprising, since the noise is 
gaussian in the data space and the reconstruction maps it into the initial space 
using a nonlinear transformation, inducing some non-gaussian features in noise in 
the process. 

\subsection{Noise in the reconstruction}

\begin{figure}
\centering\includegraphics[width=0.8\textwidth]{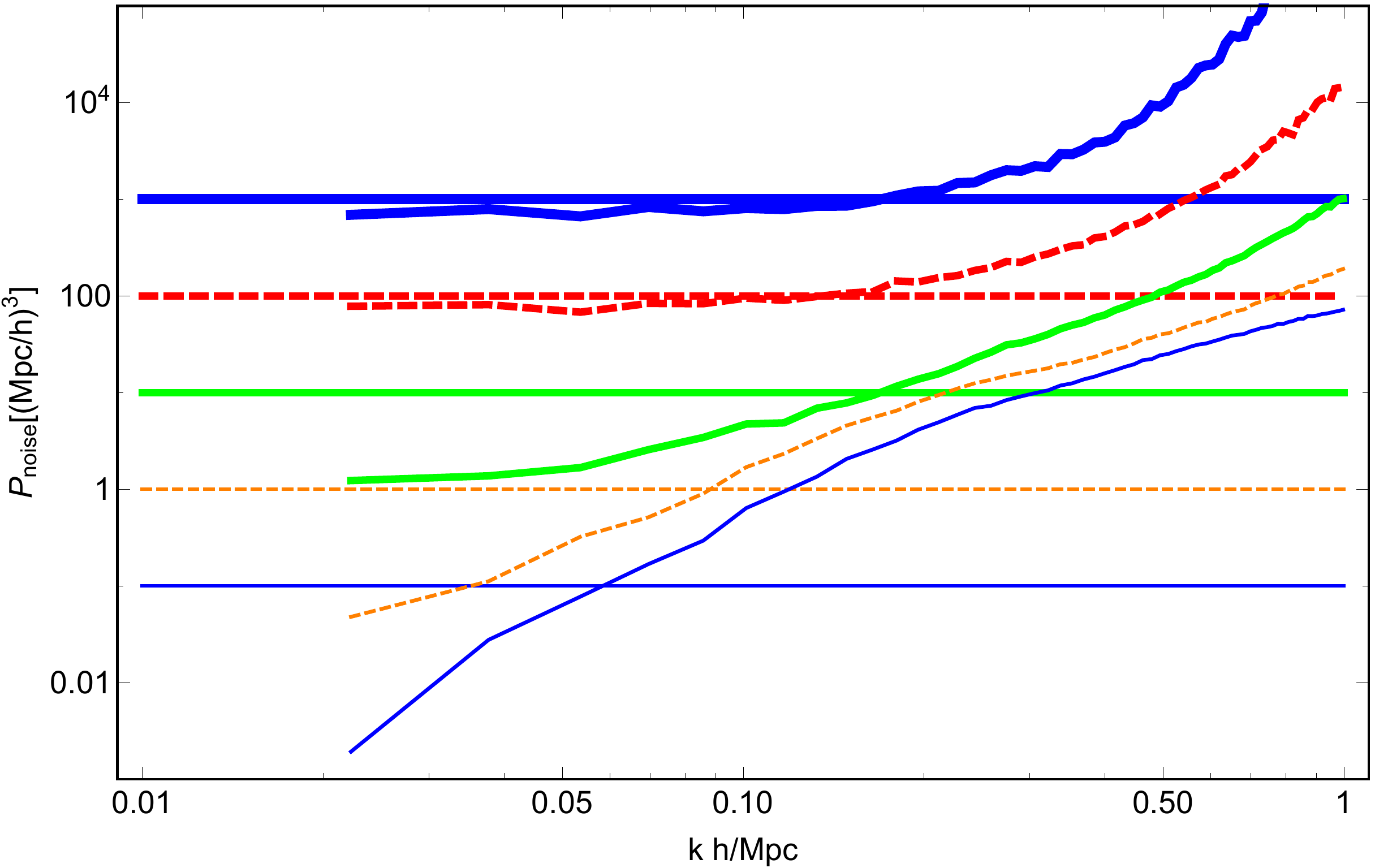}
\caption{Noise in the reconstructed linear solution. The noise is defined as 
$P_\mathrm{noise} = (1 - r_{cc}^2) P_\mathrm{truth}/r_{cc}^2$. Horizontal lines mark the linear model prediction, which is simply the level of noise in the synthetic data.}
\label{fig:recnoise}
\end{figure}

In figure \ref{fig:recnoise} we show the noise of the reconstruction, and compare it with the noise in the data. For a completely linear model, we expect the reconstruction noise to be identical to the noise in the data. The noise is defined as 
$P_\mathrm{noise} = (1 - r_{cc}^2) P_\mathrm{truth}/r_{cc}^2$: on large scales this is almost the same as
$\langle |s-\hat{s})|^2\rangle$. 

We find that the non-linearity of the model appears to reduce the noise in the reconstruction compared to the linear prediction (which the level of noise injected to the synthetic data). This is more apparent for the lower noise cases (with an order of magnitude or more reduction of the noise at the low $k$), but also evident in the more realistic high noise cases. 
We expect an effect like this from the mode coupling: if a long scale mode is coupled to a 
short scale mode then the power of the short scale mode can be informative of the long 
scale mode amplitude\citep[e.g., see][]{doi:10.1093/mnras/286.4.1023}. We can estimate a change in the amplitude of the short modes of order $R \delta_L$ with $R$ a number of order one.  Since there are about $2\times 10^6$ modes in our simulation and assuming each small scale mode 
has relative noise contribution 
of order one from the sampling variance we can  
get an error power approximately $R^2 P_{error} d^3 k /(2\pi)^3 \sim 10^{-6}$, which for $R^2=3$
corresponds to roughly $P_{error} \sim 10({\rm Mpc/h})^3$ noise power for the largest mode in our 
$400{\rm Mpc/h}$ box. What we observe in figure \ref{fig:recnoise}
for the lowest noise is even lower than that, but we suspect this is an artifact of the fact that 
we are not including scales below the Nyquist frequency in our model. 

\subsection{Towards smaller scales}
\begin{figure}
\centering
\includegraphics[width=0.7\textwidth]{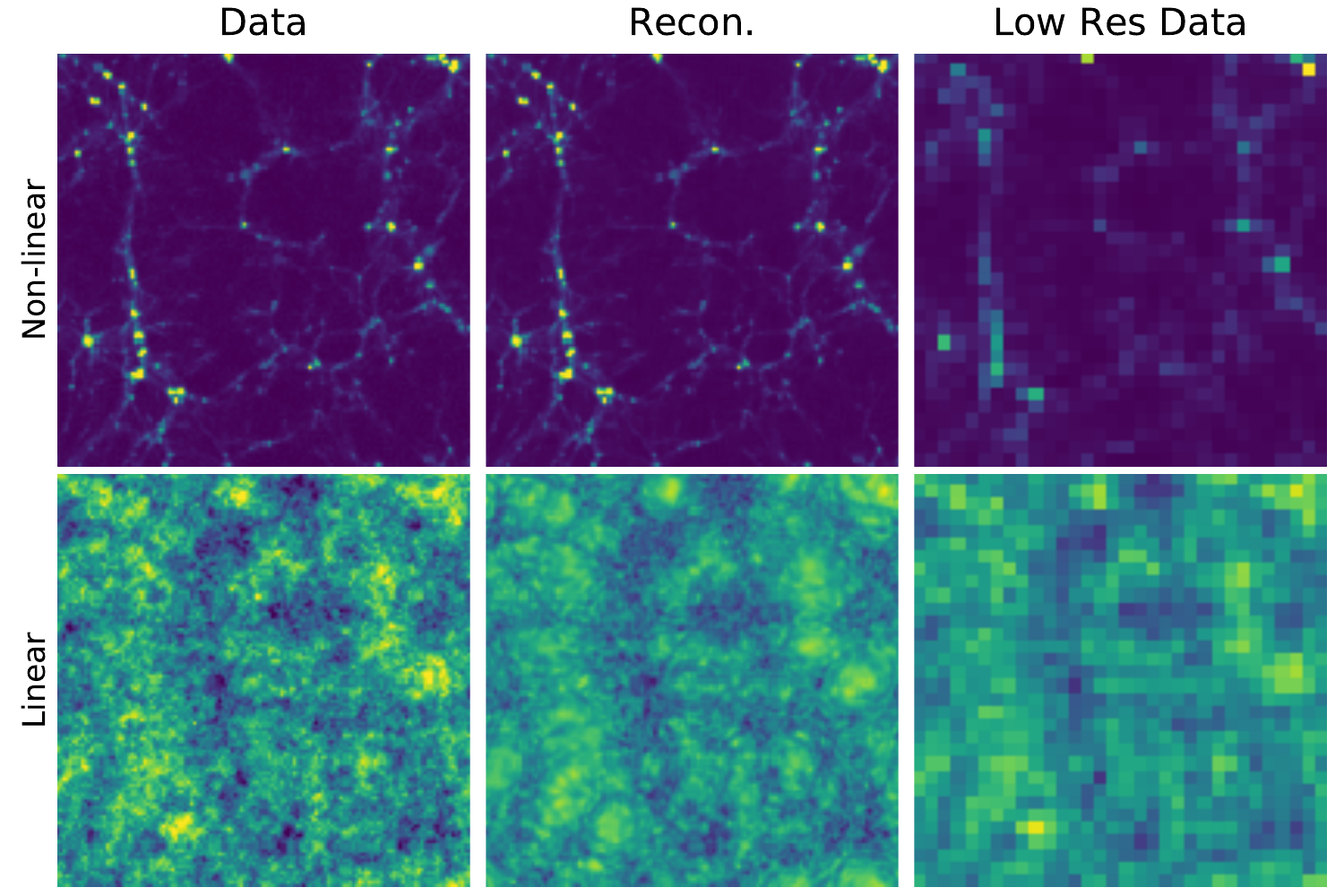}
\caption{Reconstruction of the high resolution model with N1 noise. For comparison in the third column we show the equivalent LN and NL field of the same model at the low resolution (the third column is truth, not reconstruction). The thickness of the slab is $10h^{-1}$ Mpc in all panels. Color scale fixed for each row.}
\label{fig:small-box-visual}
\end{figure}

\begin{figure}
\centering\includegraphics[width=0.3\textwidth]{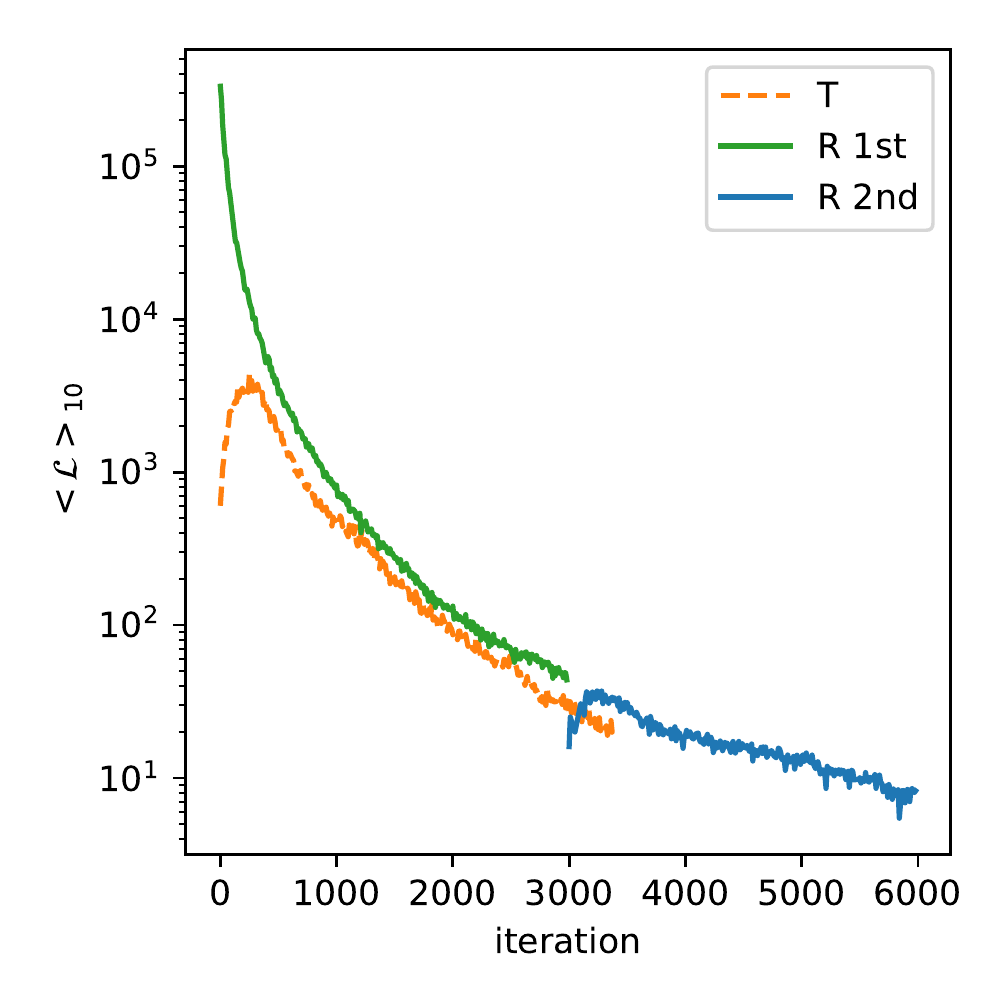}
\centering\includegraphics[width=0.63\textwidth]{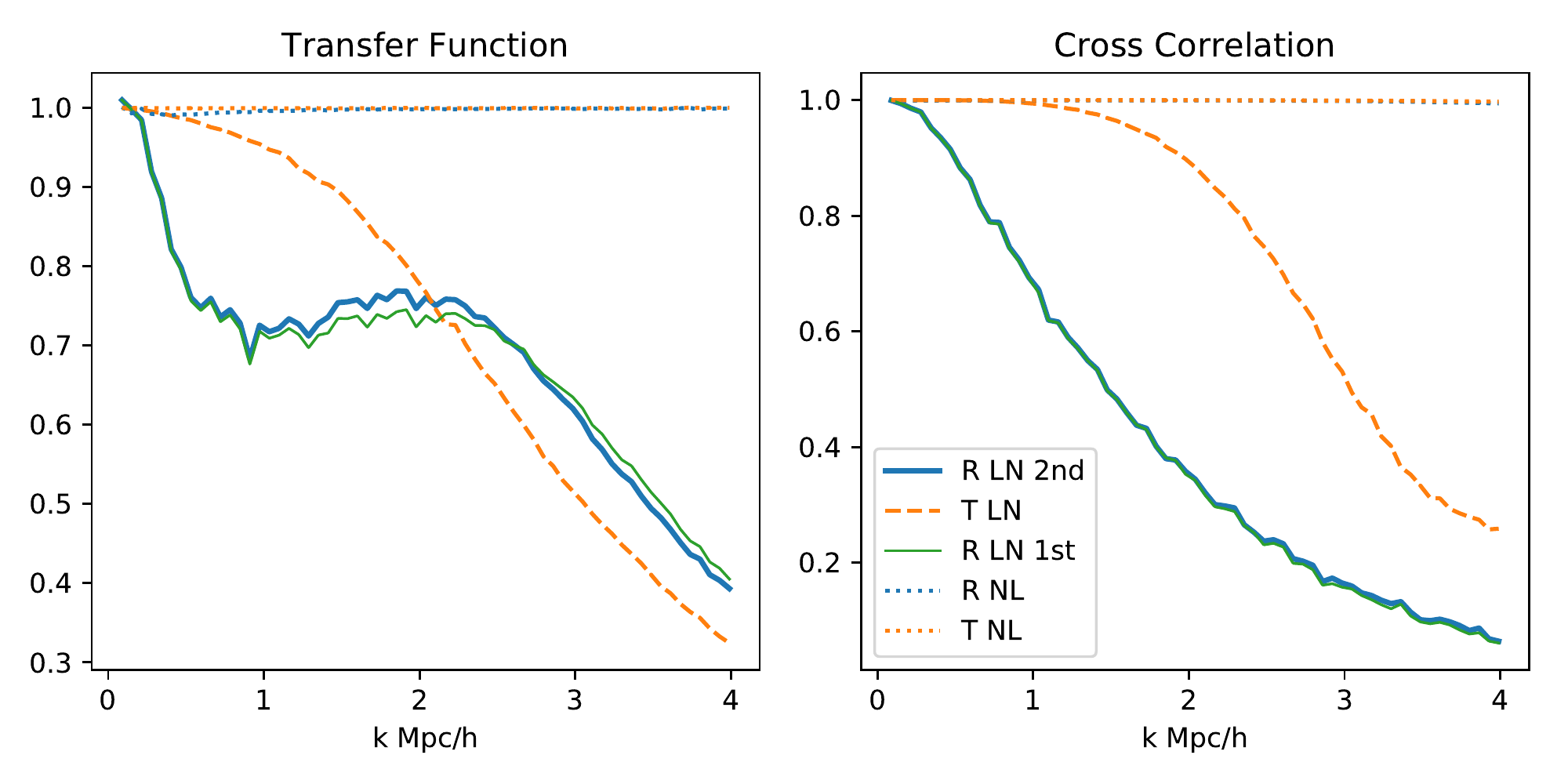}
\caption{Left panel: convergence of objective function of the high resolution model. We show the rate of descent of the objective function. We see that the convergence is much slower than the low resolution case. Middle panel: transfer function for various cases. Right panel: cross-correlation coefficient for various cases. Two different curves are starting at the truth (orange) and at a random position (blue and green, ran with two consecutive batches). The dotted lines in the upper panels are for the non-linear field and are unity for all the found solutions: we have achieved perfect 
reconstruction of the data. R-2nd is a continuation of R-1st after 3000 iterations. The L-BFGS optimizer struggles to build a reliable approximated Hessian after restarting, as evident by the slow initial descent rate in R-2nd.}
\label{fig:small-box}
\end{figure}

We investigated an even more non-linear case by shrinking the size of the simulation box from 400 Mpc/h to 100 Mpc/h, keeping resolution at $128^3$. We use the lowest noise model N1 for this 
test, since higher noise levels send small scale power to zero anyways. 
We also 
keep all other aspects of the data and model the same. The high resolution model resolves more non-linear structures than the low resolution model, as seen in Figure \ref{fig:small-box-visual}. 

In Figure \ref {fig:small-box-visual} and \ref{fig:small-box} we show the reconstruction and convergence of the high resolution case. The convergence of the high resolution model is noticeably slower: at 3000 iterations, starting from truth and starting from random are both observing a descent rate of a few times $10^{-5}$. We continued the random starting point case with an additional 3000 iterations in a second batch (R 2nd). We find that the effect of these additional iterations on the transfer function and correlation coefficient was small, up to a few percent at $k > 1$ h/Mpc in the transfer function with less than one percent improvement in cross-correlation coefficient at all scales. The final objective function of the random starting point after two optimization batches is close to 1.22 per dof, comparing to that of the truth starting point of 0.99 per dof after a single batch. In this case the random starting point minimum is thus very far from the global 
minimum, which can also be seen from the comparison of the transfer functions and 
correlation coefficients of the two starting point cases. However, in both cases we have achieved a 
nearly perfect nonlinear reconstruction, as can be seen from TF and CC being unity over entire 
range of $k$ (dotted lines). So we have high quality solutions even though the loss functions 
are very different. 

In terms of initial modes the picture is very different. The two starting points lead to a 
very different solution in terms of both TF and CC. 
Note that even our best solution has initial CC and TF less than unity on small scales, likely 
a consequence of finite noise, although we cannot exclude the possibility of 
a wrong solution being the ``global" minimum of the loss function (this is discussed further 
in the next section). The shape of TF is strange and suggestive that the solution is stuck at 
some local minimum far from the global one. This is also in line with the very poor reduced
loss function value of 1.22 per dof. It is possible that a stochastic gradient 
descent or an annealing method are able to get around this minimum, specially if it is shallow
(we could also be stuck on a saddle point with zero gradient). 

The results suggests that as the model becomes more non-linear, the reconstruction solutions obtained from the random starting point and truth starting point become more disconnected, 
and we are unable to reach the ``global" minimum with the random starting point. We should 
emphasize however that the nonlinear solution is very good in both cases, as both TF 
and CC are unity across the entire range of scales. The optimization procedure thus found 
a high quality solution in terms of the nonlinear reconstruction 
even with the random starting point, and it is only a solution with a high loss function 
value relative to the global minimum because the noise is so small. This suggests we have 
non-injective mappings, with many different initial conditions leading 
to the same final configuration. 

An important question is how the local minimum found when starting from random 
can be used for the power spectrum reconstruction. While it is considerably lower than 
the global minimum in terms of correlation coefficient and transfer function (at least for $k<2{\rm h/Mpc}$), it is reproducible, in the sense that two different starting points 
lead to the same solution, and one can correct the transfer function effects using 
a random realization. Because this is not a global minimum one must be careful in 
using the Fisher matrix construction of \cite{SeljakEtAl17} around the found solution 
and interpret it as the covariance matrix: more work is needed to establish this connection. 

It should be emphasized however that the noise level N1 is so unrealistically low that it will probably never be achieved in real data. A more realistic case is with noise level of N4 or 
N5. In this case TF and CC are already nearly zero at $k=1{\rm h/Mpc}$ and one would not 
expect that doing a higher resolution analysis would gain us anything. We were unable to 
verify this statement numerically due to the difficulty of convergence with the current L-BFGS optimizer, so we leave it to future work.

\section{1-d Zeldovich model}
\label{sec3}
In this section we explore the nature of nonlinear solutions further by exploring a simple one dimensional toy model. We will assume that we have $N_p$ particles moving on a periodic line of unit length with the dynamics given by the Zel'dovich approximation. As a result particles move in a straight line from the initial to the final conditions. The position of particle $i$, $x_i$ is given by:
\begin{equation}
x_i= q_i + a(t) \psi_i, 
\end{equation}
where $q_i$ is the initial position, $a(t)$ is the scale factor and $\psi_i$ is a Gaussian random variable which is the gradient of a potential. The index $i$ denotes the ordering in the final conditions. At the final time $a(t)=1$. 
This is a trivial example but we think it may capture some of what is going on in the full simulations in terms of reconstructions before and after shell crossings. 

We assume that at the final time we know the positions of particles or equivalently we know the density field. Thus we have a collection of final positions $\{x_i\}$ but we do not know where the particles originally come from, the $\{q_i\}$. Clearly with this simple dynamics there are many possible $\{q_i\}$ because we can always compensate and exchange in the position of two particles in the initial conditions with a corresponding change in $\psi$: the mapping is non-injective. Contrary to our full problem, here the trivial dynamics makes it easy to see that there are $N_p !$  solutions and because we are dealing with a one dimensional problem the constraint that the displacements have to be the gradient of a potential is always satisfied. Presumably in the full problem, both the more complicated dynamics and the gradient constraint will reduce the number of solutions. 

As argued in this toy problem there are plenty of solutions that reproduce the same final density in terms of the log likelihood (residual). So the 
question is how does the prior term select among these different solutions. 
As a concrete example we will assume the power spectrum of the divergence of the displacement $\partial \psi$, $P_{\partial\psi}$, is a power law and will take
\begin{equation}
{k \frac{P_{\partial\psi}}{\pi}} = ({\frac{k}{k_0}})^\gamma.
\end{equation}
In our example we generate realizations with 2048 particles and we take $k_0 =k_{Nyq}/10$ to be one tenth of the Nyquist wavenumber $k_{Nyq}$. In units of the fundamental mode $k_f= 2\pi$ then $k_{Nyq}/k_f= N_p/2 $ and $k_0/k_f = N_p/20$ and we will take $\gamma=1.2$. The most relevant parts of these choices is that $({k/ k_0})^\gamma$ is bigger than one on small scales ({\it ie.} that there is shell crossing) and that it is a growing function of $k$. 

Among all the possible solutions for the initial positions of the particles, there is a simple one where one assumes that particles have not crossed ({\it i.e.} there is no shell crossing). This solution preserves the final ordering of the particles in the line and just distributes them uniformly to get their initial conditions ($q_i= (i-1) /N_p$) \cite{2016arXiv160907041Z}. The associated displacements are given by the difference between final and initial conditions. This reconstruction is similar to the ones implemented in 3D in \cite{ZhuEtAl17,SchmittfullEtAl17}. 

\begin{figure}
\includegraphics[width=\textwidth]{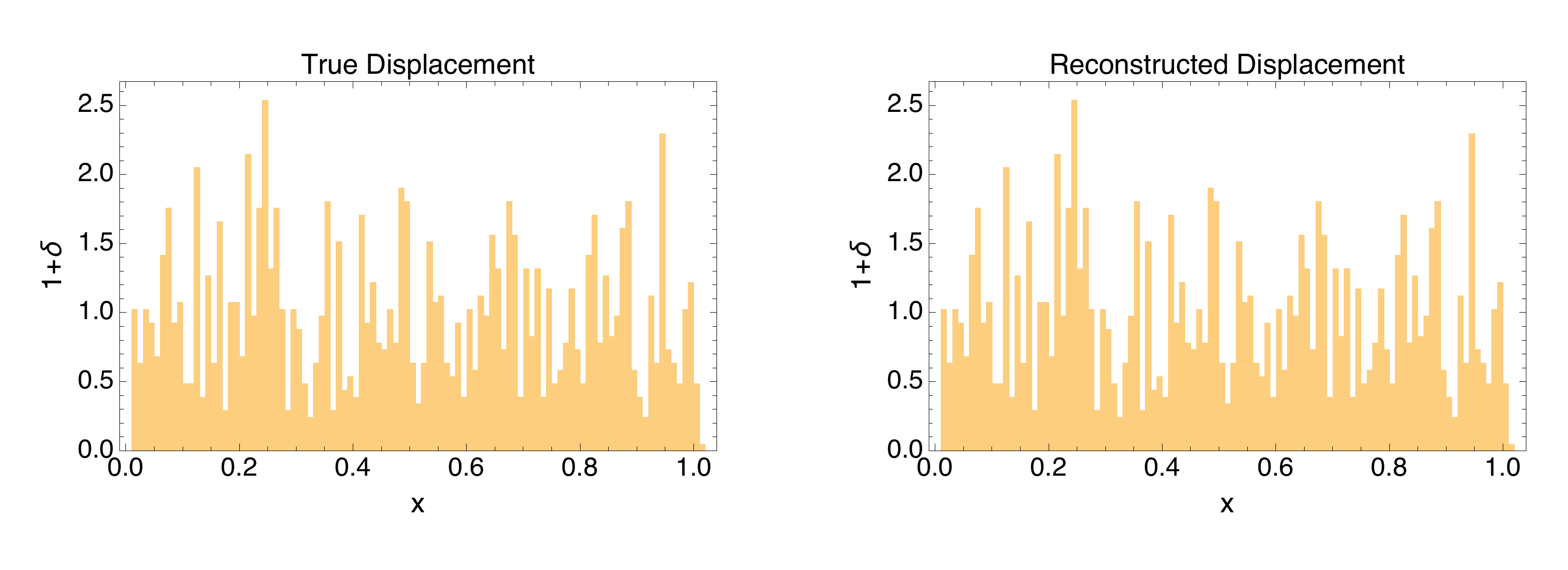}
\caption{Density field of a single representative realization, final density and the density evolved with the reconstructed displacement are by construction identical. }
\label{fig:toy-densities}
\end{figure}

\begin{figure}
\includegraphics[width=\textwidth]{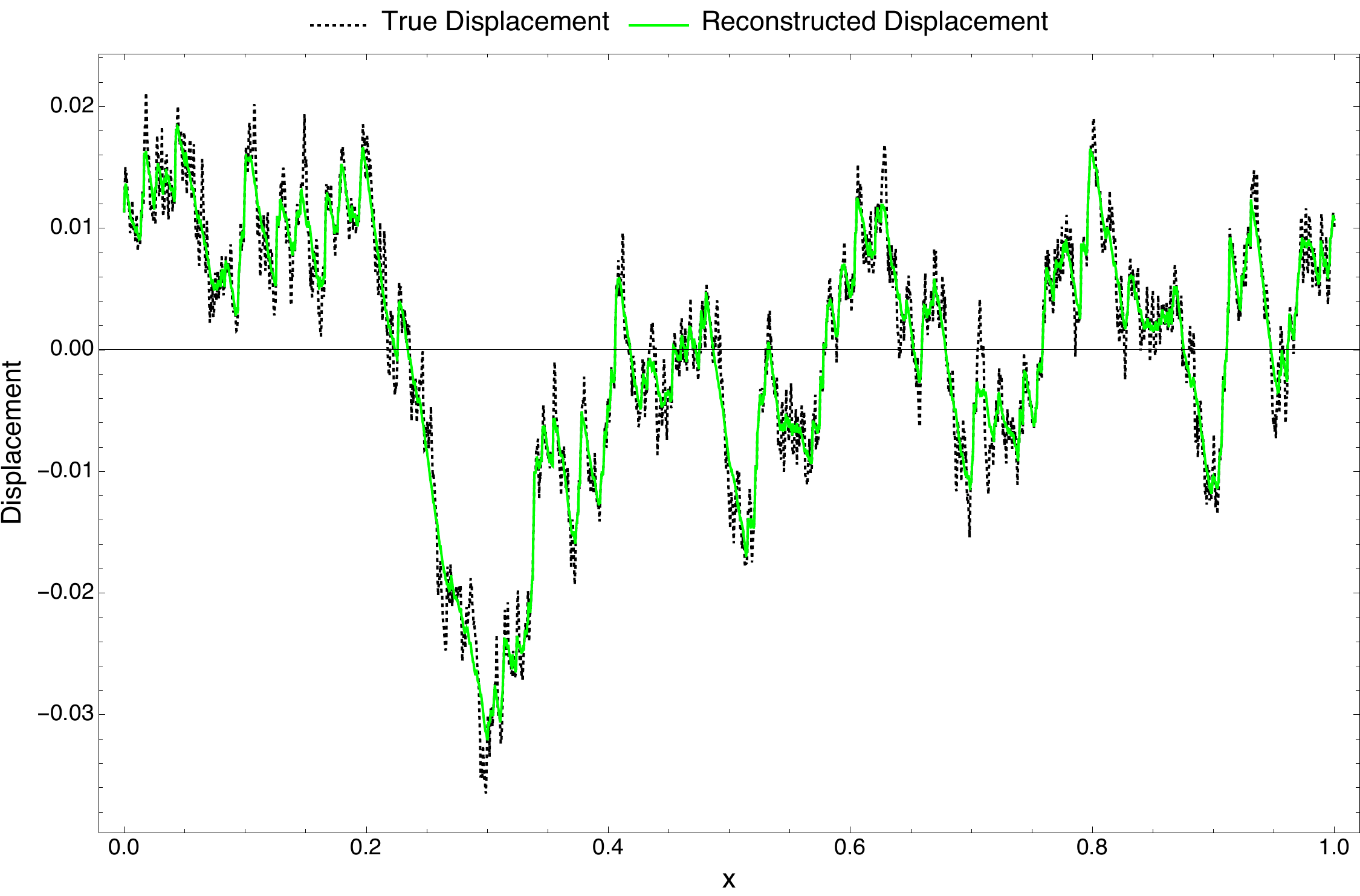}
\caption{Reconstructed and true displacements. the reconstructed field is clearly smoother.}
\label{fig:toy-displacements}
\end{figure}

In figure \ref{fig:toy-densities} we show a sample realization of the final density field, as well as the density field produced by the reconstructed displacements. By construction the two are identical even if the displacement fields are not the same. In figure \ref{fig:toy-displacements} we show both the true displacements as well as the reconstructed displacements. It is very clear from this plot that the reconstructed displacement field is significantly smoother. This is a direct result of the demand that there be no crossing between the particles. 

\begin{figure}
\includegraphics[width=\textwidth]{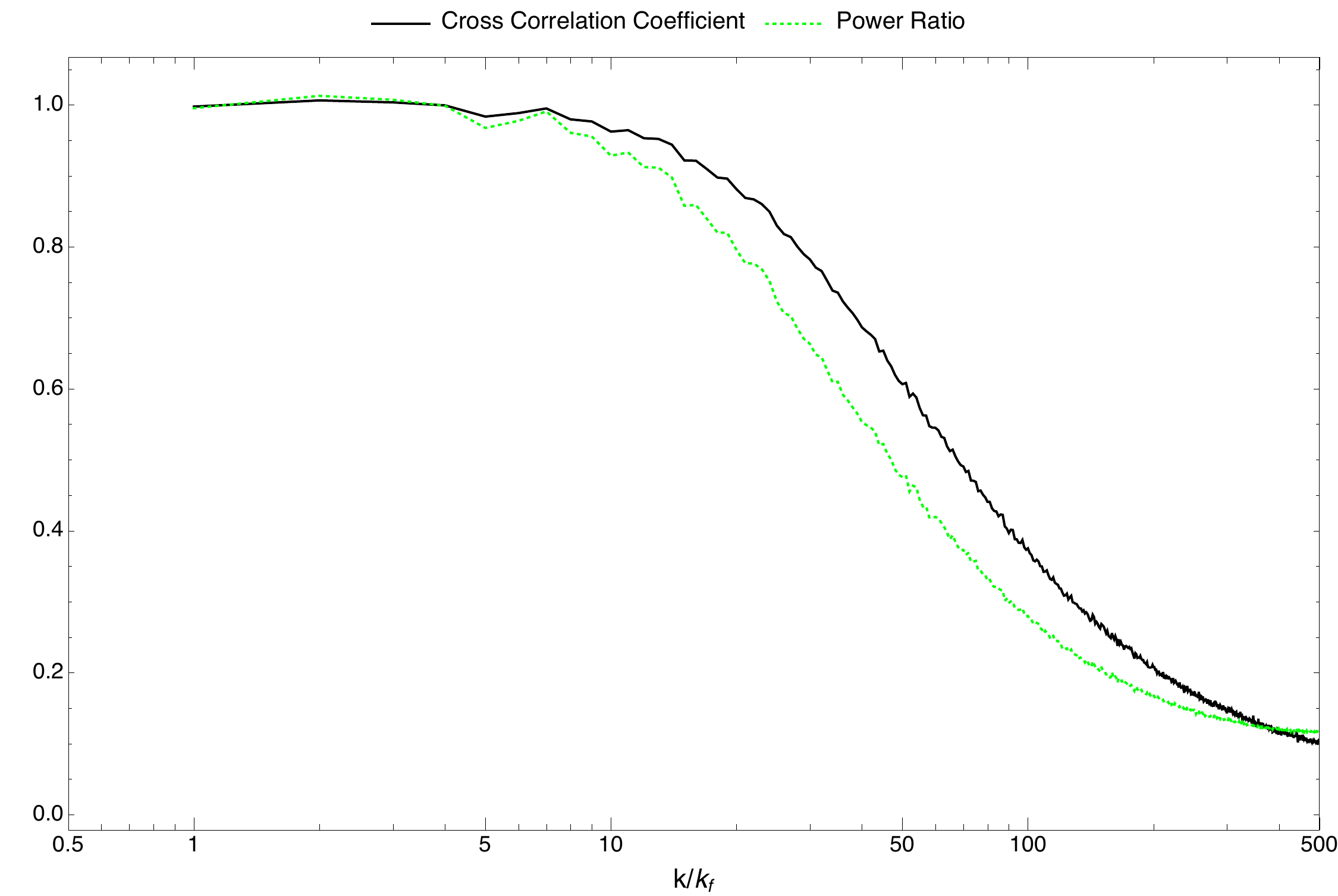}
\caption{Ratio of the power spectra of the reconstructed to the true displacement and cross correlation between the two fields. In both cases we show results averaged over $10^4$ realizations.}
\label{fig:toy-cross}
\end{figure}

The power spectra of the reconstructed and true displacement fields are also different, with the reconstructed displacement having smaller power on small scales. We show the ratio of power spectra in figure \ref{fig:toy-cross}. At the scale $k_0$ the reconstructed power is about 30\% of the true one. In this figure we also show the correlation coefficient between the reconstructed and true initial conditions. We see here a loss of correlation as one goes to small scales, similar to the results 
of full 3d analysis. Even though we are finding displacement fields that lead to the same final density on all scales, we are not finding the correct initial conditions for the small scale modes. It is obvious that the demand that there be no shell crossing leads us to an incorrect reconstruction.  

\begin{figure}
\includegraphics[width=\textwidth]{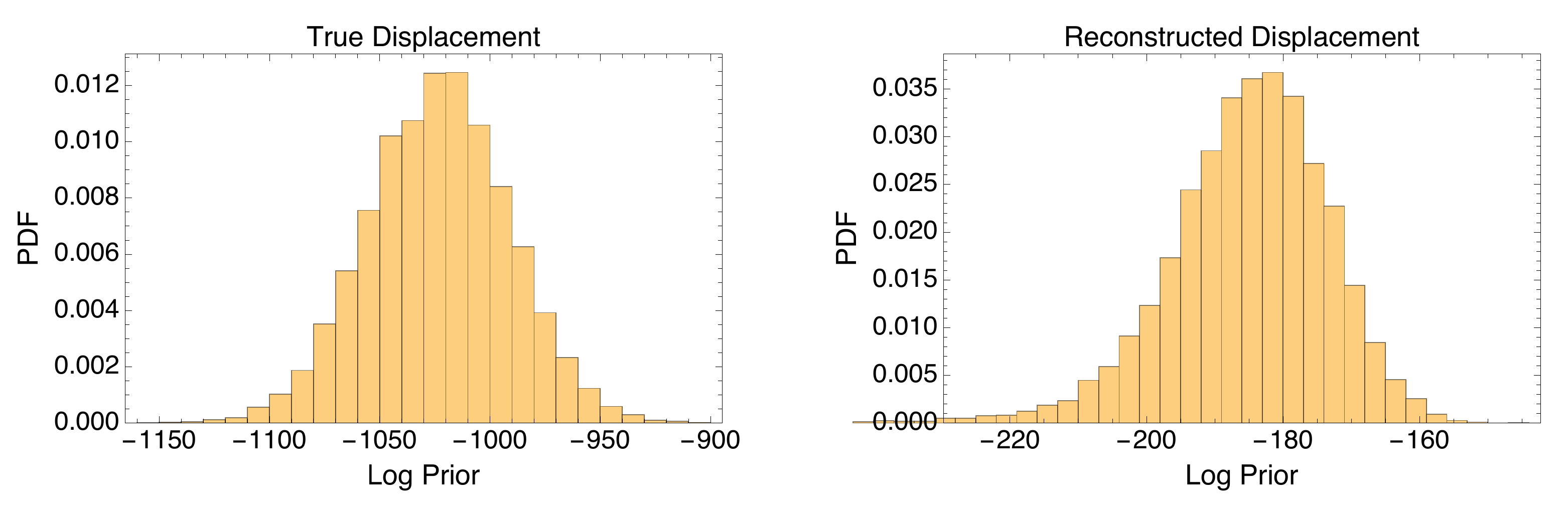}
\caption{Distribution of the logarithm of the prior over $10^4$ realizations for both the true displacements and the reconstructed ones. The decrease lack of small scale power in the reconstructed field leads to a substantially larger probability.}
\label{fig:toy-logP}
\end{figure}

In terms of comparing these solutions to the data, both the true displacement and the reconstructed one produce the same final density and thus lead to the same likelihood of the data given the model. They differ however in their prior probabilities. 
It is important that because the reconstructed field has smaller power on small scales it will always have a larger prior probability. In figure \ref{fig:toy-logP} we show a histogram of the log of the prior (as defined in equation \ref{eq:obj}) over our $10^4$ realization. 
This implies that if one were to just look for the highest probability reconstruction one would pick the wrong answer. The problem would likely disappear if one could sample over the different solutions, as the increase in the probability of the solutions with smaller power may be compensated by their smaller phase space volume, but this requires finding all the solutions which in 
very high dimensions suffers from the curse of dimensionality and is in general impossible. 

We see there are some analogies between the toy 1-d Zeldovich example and 
the full 3-d solutions described in previous sections. 
First of all, in both cases the reconstructed solution 
is reduced (in TF and CC) relative to truth. However, in 3-d this could be because of 
the presence of a small, but finite, noise. This is suggested by the fact that 
even the solution starting at truth moves away from this solution to one with 
reduced TF and CC. 
We were unable to reduce the noise 
further in our 3-d solutions, since the gradient was becoming numerically unstable, so we are 
unable to verify that there is a reduction of power due to shell crossings in the 
limit of zero noise in 3-d.

Another similarity is that in 3-d the solution starting at the truth, while best in terms of 
residuals, is worst in terms of the prior: this is also similar to the 1-d case, 
where the reconstruction finds the solution among those with zero 
residual that has the lowest prior, but this solution is not the correct one. 
It seems plausible that as we push to 
smaller scales, with more shell crossings, this effect will get stronger and the 
reconstructed solutions starting at random or zero 
will deviate from the truth more and more even in the 
low noise regime. 

Third similarity is that both in 1-d and in 3-d we seem to have evidence of 
non-injective mappings, with several different initial conditions leading to nearly 
identical final state. The difference is that in 1-d the true solution is not at the 
global minimum, while in 3-d we have no evidence that it is not. 

There are however several issues that prevent us from making more conclusive 
statements. One is that optimization only finds a local minimum and we have 
no proof that we have found the global minimum. In the 1-d toy model the 
dynamics is sufficiently simplified that the (many different)
solutions can be simply written down in analytic form and we were able to 
identify the global minimum solution and show that it does not correspond 
to the true solution. In the full 3-d we were not able to show this: solutions 
starting at truth have lower loss function value than those starting at 
random or zero. We do not know whether this is because we are working on relatively 
large scales where shell crossings do not play a role, or whether this is 
because the gradient constraint in 3-d reduces the number of solutions. 
However, these issues are mostly of academic interest, as for realistic noise 
levels noise sends modes at high $k$ to zero anyways. 

\section{Discussion and conclusions}
\label{sec4}

In this paper we explore some of the conceptual and practical issues regarding 
the posterior surface of the initial condition reconstruction 
problem in LSS. Specifically, we ask 
whether the reconstruction of initial linear density field is unique as we vary 
both noise and resolution. For this purpose we use finite step N-body 
simulations (using the FastPM scheme) that have been shown to agree well with the
full N-body simulations over the range of scales we have applied to here.
Our approach uses optimization: we write the posterior and 
maximize it with respect to initial linear modes. 
For high noise levels the small scale structure cannot be
reconstructed and is instead sent to zero, 
as is expected due to the corrupting power of noise. In the high resolution low noise case we see some 
evidence of non-convexity in the posterior surface, in that starting from the 
truth converges to a different solution than starting from random or zero starting point, 
but for the more realistic noise and lower resolution case these effects are relatively
small. 

The quality of solutions improves overall as we reduce the noise. We see this both in terms 
of increase in CC and TF at high k, as well as in terms of reduced noise of reconstructed modes 
at low k. For lower noise levels we observe that the error is dramatically reduced at low k. 
We interpret this with the nonlinear mode coupling: small scale modes are all coupled to the large 
scale modes and can provide information about them that is lower than the noise in the large 
scale mode itself. Of course this information is of limited use for the power spectrum which 
is sampling variance limited on large scales, but probing other parameters 
such as primordial non-Gaussianity may benefit from this noise reduction. 

This paper explores the posterior surface of the modes. As mention in the introduction, 
we do not actually care about the modes per se, but what they can tell us about the 
power spectrum reconstruction.
For the most part any realistic starting point leads to the same 
quality solution. So from this point of 
view our MAP solutions, even if not unique, are all of comparable quality. 
Specifically, this means that one can use a random simulation to calibrate these
solutions and correct for the missing power, if the initial power spectrum reconstruction 
is the final outcome of the analysis \cite{SeljakEtAl17}. We see no evidence that this 
correction is realization dependent for noise levels we can achieve in the current future 
data.


At the lowest noise results change somewhat as we increase 
the resolution by a factor of four: in this case we have clearer evidence of a
multi-modal posterior, with solutions found by optimization starting at random 
being very far from the ``global" maximum posterior solution found when optimization starts at truth. However, even in this case the solutions are all very good in terms of the 
final nonlinear reconstruction, so there are many different initial conditions that 
lead to nearly the same final configuration. The mapping induced by nonlinear 
gravity is thus non-injective, at least at a finite resolution we work with here. 
If the initial power spectrum is the ultimate goal its bias
can be calibrated, but may lead to a larger variance than at the global maximum. In real data we
have no way of obtaining the global maximum in the high number of dimensions 
explored here. 

We use 1-d Zel'dovich case to show analytically the non-injective nature of the 
mappings. We argue that once the 
shell crossings occur it is not even guaranteed that the global maximum 
solution will be the true solution even in the low noise limit. Once 
the forward mapping becomes non-injective using just the final 
position there is no unique way to find the original configuration. One would hope 
that the prior would allow one to choose the true solution among the many different 
solutions that match the data, but using this 1-d Zeldovich model we show that this 
expectation is not met, and the prior chooses instead the solution without 
shell crossings, which in general is not the true one. 

In our 3-d examples we did not find any evidence for the global maximum to be further 
away from the truth than some other maximum. 
Starting 
optimization at the truth always led to a lower loss function than starting at any 
random starting point. This could be because we have not yet reached the 
shell crossing scales, even at $k=4{\rm h/Mpc}$, or it could be that the 3-d dynamics 
is more restrictive, or 
it could be that our solutions when not starting at truth 
are stuck at local maxima (or saddle 
points) far from the global one. 
Some evidence of this is that when going to smaller scales 
our optimizer slowed down dramatically, 
suggesting the posterior surface is complicated and likely multi-modal on small scales. 
It is possible that better global optimization methods that use annealing or some other 
method may be able to significantly 
reduce the loss function value in the low noise regime, but any global optimization methods suffer from the curse of dimensionality and we are not optimistic that one can make much progress in 
the general case. One can for example try different starting points, but we have already seen 
that any realistic initial starting point leads to nearly the same solution: it is just that 
this is not the solution we get when starting from an unrealistic starting point, the truth. 

Because of this we also do not believe MCMC sampling methods can do better in very high dimensions: 
a typical MCMC such as HMC will descend to the nearest local posterior maximum (burn-in phase), 
and then sample around it such that the local posterior surface is explored. In the limit of 
very low noise the sampler will not move away from the local maximum, and it will 
not find the existence of other local maxima. Moreover, the sampler will be fooled into 
thinking that the local maximum is very deep 
close to a Dirac delta function, and it will 
conclude that there is no error in the mode reconstruction. Since the sampler has no information 
about the existence of the other local maxima, it will also conclude that it has measured the 
power spectrum at the sampling variance limit, and it will not correct for the bias that arises 
from the fact that these modes are not the true modes. Using MCMC blindly is more prone to 
a bias than using the methods of \cite{SeljakEtAl17}, where the bias is explicitly determined
using a simulation. It may be possible to develop annealing style sampling methods where 
existence of multiple maxima is probed during the burn-in phase, and then each is 
sampled, but it is far from guaranteed 
that this would be competitive, and in any case no such method has been developed yet. 
Optimization methods can also be enhanced with some randomness such as annealing so 
that they can move out of a shallow local minimum, and it is worth pursuing these
methods both in MCMC and in optimization context to see if they improve the quality of 
solutions. 

While the question of multi-modality remains unanswered, it is of mostly academic interest: for 
realistic noise levels that can be achieved in current and future LSS surveys 
small scale modes will not be reconstructed because of the noise, not because of the
shell crossing degeneracies. In this case the differences between the 
different maxima in the posterior are negligible when realistic initial conditions 
are used. 
The proposed LSS analysis
program of \cite{SeljakEtAl17}
 is to use initial density reconstruction method as a starting 
point for analytic marginalization over the latent variables, leading to the 
initial power spectrum estimator as a summary statistic. 
The method studied in this paper appears to be  
a viable strategy that can be both computationally 
feasible and nearly optimal for realistic noise levels of LSS surveys. It is possible that 
at the highest noise levels even simplified dynamics based on low order perturbation theory might be good enough: the dark matter density computed in this way has a higher correlation coefficient with the density computed using the true dynamics than our full reconstruction using N-body based on noisy data. However, correlation coefficient does not contain all the information and if the reconstructed
field from simplified dynamics is non-gaussian then higher order correlations will also matter. 
It is only 
for the extremely low noise levels that the multi-modal nature of posterior surface 
becomes problematic, in that the global maximum cannot be reached by optimization 
or sampling methods, leading to a loss of information because of that. 

\vspace{.1in}
\textbf{Acknowledgment} 
The analysis pipeline of this work is built with nbodykit \cite{nick_hand_2017_1051244}. The majority of the computation were performed on NERSC computing facilities Edison and Cori, billed under the cosmosim and m3058 repository. We acknowledge support of NASA grant NNX15AL17G. 
M.Z. is supported by NSF
grants AST-1409709 and PHY-1521097, the Canadian Institute for Advanced Research
(CIFAR) program on Gravity and the Extreme Universe and the Simons Foundation Modern Inflationary Cosmology initiative.

\appendix
\section{Adiabatic Optimization}
\label{appa}
The number of modes scales as $k^3$ and there are far fewer large scale modes 
than small scale modes, as the number density of modes per uniform $k$-bin increases as $k^2$. As a result, the objective function is more sensitive to the small scale power than the large scale power, causing slow  convergence of the large scale power. In the quasi-linear regime, gaussian smoothing eliminates the sensitivity on small scales, allowing the optimizer to focus on the large scale modes. 
Once these have converged they are no longer changed as we adiabatically add back 
small scale power, and the entire scheme enhances the convergence of the optimizer. 

We define a family of objective functions

\[
\mathcal{L}^{\hat{G}}({\bi s}) = \sum_k \frac{|s_k|^2}{S(k)} + \sum_i \left(\frac{\hat{G} [d_i-f_i(\bi{s})] }{\sigma}\right)^2, 
\]
where $\hat{G}$ is a Gaussian smoothing window. During the optimization, we adiabatically decrease the size of the Gaussian smoothing window from 16 Mpc/h to zero. At each smoothing scale above zero, we terminate the optimization when the objective function decreases less than 1\% of the total number of modes. Typically a total of less than 100 iterations with nonzero smoothing are performed before we switch to the final objective function without 
smoothing (as defined in \ref{eq:obj}).

\section{Automated Differentiation}

Optimization in a very high number of dimensions is only possible if the gradient 
of the loss function with respect to optimization parameters (in our case modes $s_k$) 
can be evaluated. Gradient of the prior is trivial, but for the gradient of the 
log likelihood one needs the derivative of the data model $df_i/ds_k$. 

The chain rule relates the Jacobian of a model $f(x) = z(y(x))$ to a product of Jacobians of known operators ($z(\cdot)$ and $y(\cdot)$ in this case),

\be
J^{zx}_{ij} = \frac{\partial z_i}{\partial x_j} = \sum_k \frac{\partial z_i}{\partial y_k} \frac{\partial y_k} {\partial x_j} = \sum_k {\Large J}^{zy}_{ik} {\Large J}^{yx}_{kj}.
\ee

The topic of automated differentiation is covered in several text books, e.g. \cite{NocedalWright06}. In short, an important observation on the chain rule above leads to automated differentiation. 
We notice that a vector can be applied on either sides of the full Jacobian. We will call the left application the vector-Jacobian product (vjp), $\sum_i  \mathrm{v}_i J^{zx}_{ij}$, and the right appliation the Jacobian-vector product (jvp) $\sum_k J^{zx}_{ij} \mathrm{v}_j$. 

For the vector Jacobian product, we notice

\be
\sum_i \mathrm{v}_i J^{zx}_{ij} = \sum_k 
\left( \sum_i \mathrm{v}_i {\Large J}^{zy}_{ik} \right) {\Large J}^{yx}_{kj}
\ee

Therefore, if we know the vjp of $z(\cdot)$ and $y(\cdot)$, then we can compute the vjp of $f(\cdot)$. This mode of automated differentiation is also known as the back-propagation. Similarly, forward-propagation reduces the chain rule into a sequence of Jacobian-vector product (jvp) operators. For a detailed and practical description, we refer to a recent implementation of automated differentiation described in \cite{2016PhDT.......317M}. 

Here we point out an special case computing the gradient of the objective function $\mathcal{L}(s)$ in equation \ref{eq:obj}. Because $\mathcal{L}(s)$ is a scalar valued function, the Jacobian and the gradient are the same. Therefore, we can apply the backward propagation rule with a starting vector $v=(1,)$ that is 1 dimensional to compute the gradient.

\be
\mathbf{g}_k =  \mathbf{v}_{0} J^{\mathcal{L},s}_{0k}.
\ee

The function $\mathcal{L}$ is a summation of squares. The complication comes in the forward model $f(s)$. $f(s)$ is a PDE solver in the form of a N-body simulation with many time steps.

\begin{figure}
\centering \includegraphics[width=0.7\textwidth]{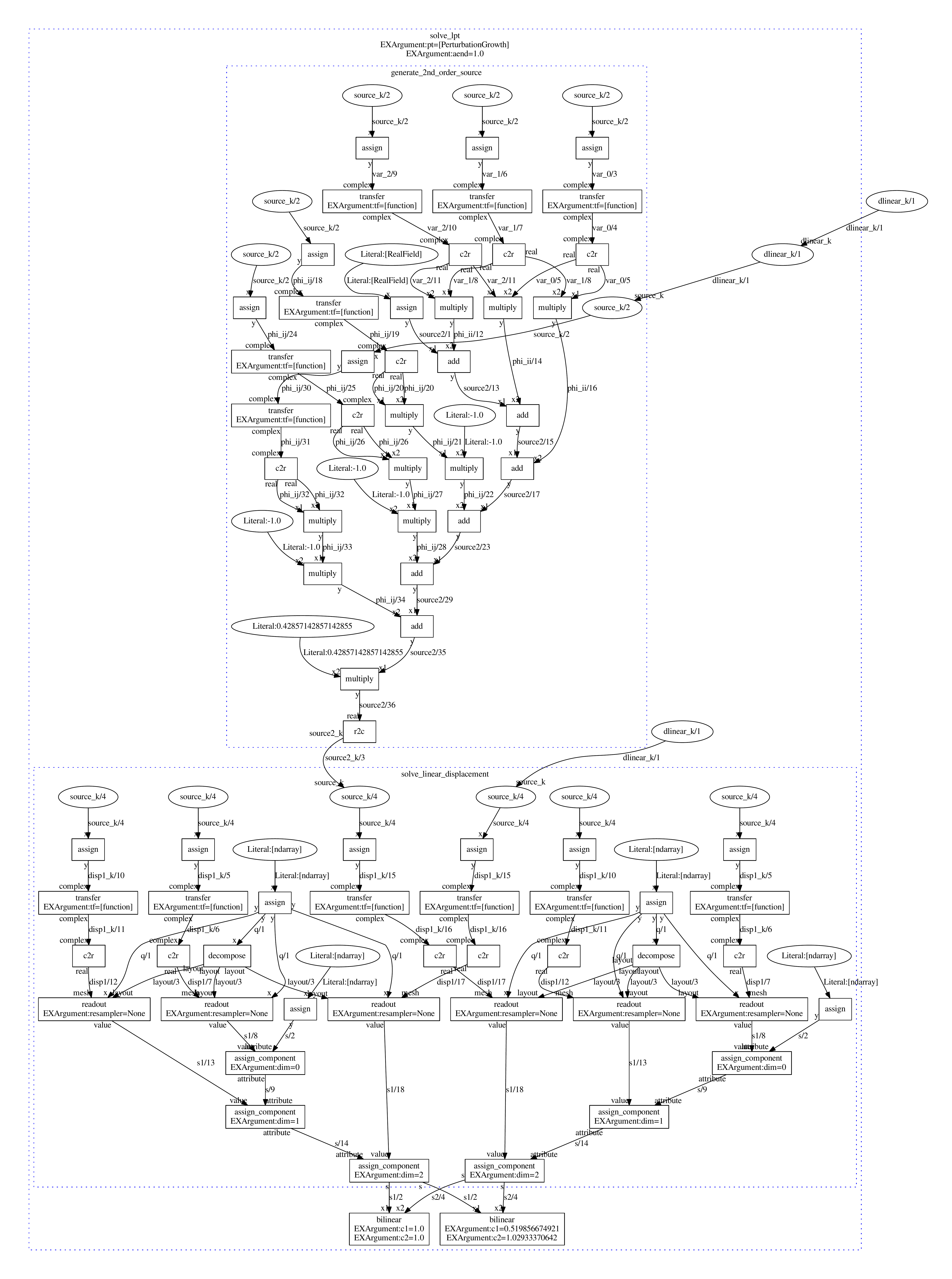}
\caption{The full computing graph of a second order LPT simulation. The purpose of this figure is to demonstrate the complexity of realistic models. The computing graph of a 10 step PM scheme is approximately 10 times larger.}
\label{fig:lpt-autodiff}
\end{figure}

The jacobian of a N-body simulation is one of those trivial but complicated problems. We illustrate the computing graph of the second order LPT solution (the zeroth step) in Figure \ref{fig:lpt-autodiff}. Although it is possible to derive manually both Jacobian products (jvp and vjp) of any model, for complicated model the endeavor quickly becomes unmanageable. The complexity of models grow very quickly as soon as computer simulations or ordinary differential equation solvers are involved. 
It also becomes tricky to recognize patterns of computable subroutines from the derived equations \citep[e.g., see][]{WangMoEtAl14,WangMoEtAl13}. 

This is exactly one of the problems automated tools are developed to aid: the process of translating a well-formed model (usually in the form of a computing graph) into its Jacobian products. Outside the field of cosmology, automated differentiation is widely used in training of numerical models (e.g. machine learning, weather forecast), and a large variety of tools have been developed.  In recent years AD tools have been developed for deep-learning toolkits (e.g. tensorflow, torch etc.). 

There are two major flavors of AD implementations: static tools based on static source code analysis (e.g. Adol-C); and dynamic tools that uses runtime graph construction with a tape (e.g. autograd). Static AD tools can either use a customized modeling language to aid the source code analysis (Stan), or to reuse the parser of the hosting programming language (e.g. tensorflow, PyMC3) via type-inheritance and operator overloading. Dynamic AD tool are also implemented via type-inheritance and operator overloading features of the hosting programming language (e.g. pytorch and autograd) as well.

The core of an automated differentiation tool consists of a graph translator that translates either a tape (sequence of operators with operands and results) or a model (just a sequence of operators) to a computable Jacobian product graph. Performance and usability are both desirable properties of the design, and some trade-offs must be made: special optimizations add complicity in the interface for declaring new operators and models. We are also concerned about the size of problems: for large models that are memory intensive, a distributed computation model must be deployed where special care must be taken for shuffling and reduction of distributed variables. 

In this work we implement a toolkit of modeling, automated differentiation and minimization in an experimental software package ABOPT (Abstract Optimizer, http://bccp.berkeley.edu/software/). The automated differentiation tool in ABOPT supports forward propagation and backward propagation of the first order derivative of a model. A model is represented as an acyclic graph, internally stored as a linearized sequence of nodes (operators), where the edges are the variables. ABOPT contains an implementation of the L-BFGS optimization algorithm.

We implemented a few frequently used linear algebra operators and their gradients following the discussion in \cite{IMM2012-03274}. In addition to regular linear algebra operators, the particle mesh simulations employed in this work rely on two resampling operators that relate a structured density field on a mesh and a unstructured density field represented by particles\citep{HockneyEastwood81}.
The particle to mesh resampling operator (paint) is
\[
  A_i(x, B) = \sum_j W(q_i - x_j) B_j,
\]
while the mesh to particle resampling operator (readout) is
\[
  B_i(x, A) = \sum_j W(x_i - q_j) A_j.
\]
In either case, $q$ marks the fixed mesh points, $x$ the coordinate of particles, and $W(r)$ is the re-sampling window function (e.g. Cloud-in-Cell, which is first order differentiable). The Jacobian of these operators are more complicated because they are functions of two vectors, the coordinates $x$ and the field $B$. 

For the particle-to-mesh resampling operator, we have
\begin{eqnarray}
  \frac{\partial A_i}{\partial x_k} &=& \sum_j 
  \frac{\partial W}{\partial r_{ij}} \frac{\partial r_{ij}} {\partial x_k} B_j \\
  \frac{\partial A_i}{\partial B_k} &=& W(q_i - x_k).
\end{eqnarray}

For the mesh-to-particle resampling operator, we have
\begin{eqnarray}
  \frac{\partial B_i}{\partial x_k} &=& \sum_j 
  \frac{\partial W}{\partial r_{ij}} \frac{\partial r_{ij}} {\partial x_k} A_j \\
  \frac{\partial B_i}{\partial A_k} &=& W(x_i - q_k).
\end{eqnarray}

The gradient of the window $W$ function is a direct product of the separation $r$ in different directions, with the same support (size of the nonzero range of the function). As a result, the Jacobian product operators can be easily implemented together with the original paint or readout operators by replacing the window with the differentiated version.

\begin{figure}
\includegraphics[width=\textwidth]{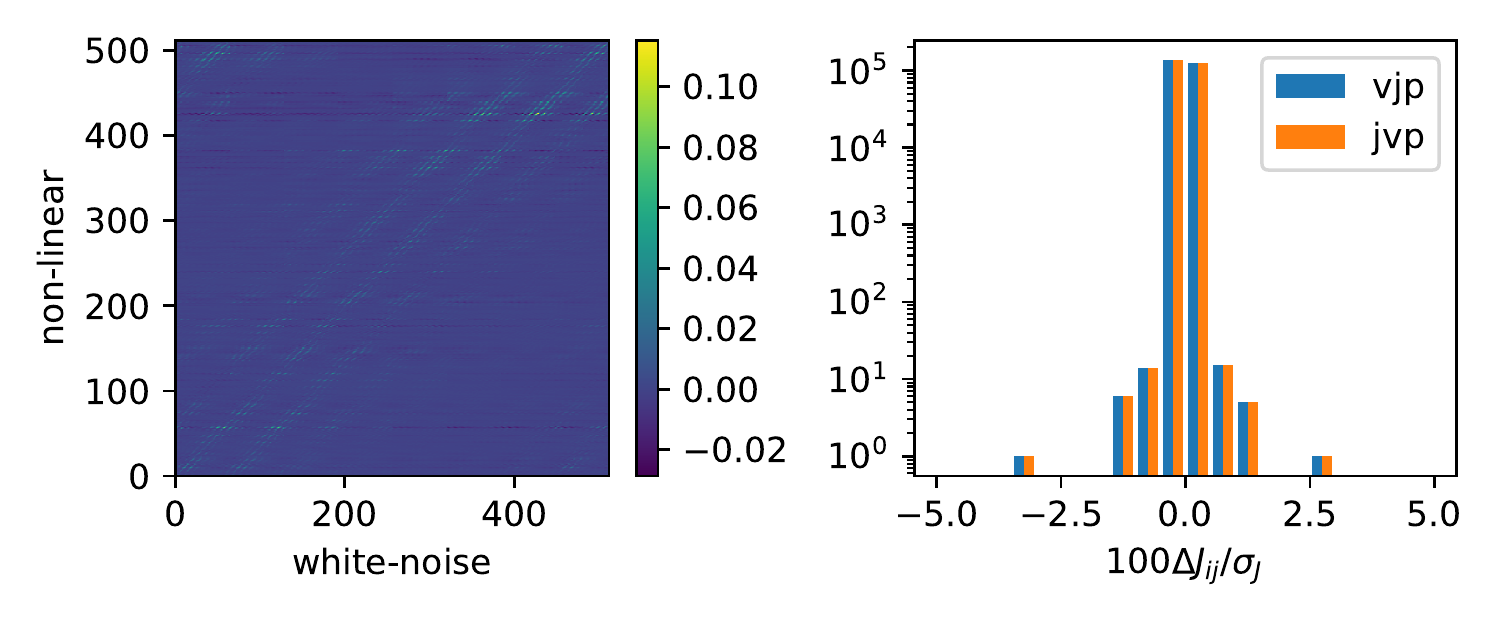}
\caption{Comparing numerical Jacobian against automated differentiation Jacobians. A simulation of $8^3$ particles in a box of 32 Mpc/h per side is performed. The initial condition is a white noise field of $8^3=512$ real numbers, draw from a Gaussian field. The non-linear density is calculated with a 10 step FastPM scheme. In the left we show an image of the full Jacobian of the model. In the right we compare the automated differentiation Jacobian against numerical Jacobian obtained from symmetric finite differentiation. The discrepancy is less than 5\% of the standard deviation of the Jacobian elements}
\label{fig:gradient-verify}
\end{figure}

As an example, we perform a numerical experiment to evaluate the correctness of our automated differentiation implementation in ABOPT. The experiment is performed on a model of $8^3$ mesh in a box of 32 Mpc/h per side, with 10 FastPM steps. The initial condition is drawn from a real Gaussian density field, $g^{\mathrm{WN}}$, consisting of 512 real numbers. The final condition is the non-linear density field $\delta^{\mathrm{NL}}$ in a $8^3$ mesh, which also consists of 512 real numbers. 

\be
\delta^{\mathrm{NL}}_i = f_i(g^{\mathrm{WN}}_j, \dots),
\ee

The forward model in the main text (equation 2.3) is equivalent to the model here up to the change of variable from the linear density field $s$ to $g^{\mathrm{WN}}$. The reason to work with $g^{\mathrm{WN}}$ is that it is easier to work with the real numbers instead of the modes because the hermitian conjugate properties of the field is automatically manifest.

We generate 512 base vectors, each of which are fed through the forward and backward propagation of the model to construct the full, 512 by 512 Jacobian matrix.

The numerical Jacobian is constructed via symmetric finite differentiation with two impulses of $\pm 10^{-3}$ on each variable in the white noise field. The difference of the two non-linear density field forms the corresponding Jacobian element. We find that the forward and backward propagation give nearly identical Jacobian elements (relative difference $<10^{-7}$). The numerical Jacobian agrees very well with the automated Jacobian, with two parameters that are a three percent off and the rest within one percent. These results are shown in figure \ref{fig:gradient-verify}.

\bibliographystyle{JHEP}
\bibliography{cosmo,main}

\end{document}